\documentclass[journal]{IEEEtran}
\ifCLASSINFOpdf
\else
\fi

\usepackage{url}

\usepackage{amsmath}
\usepackage{graphicx}
\usepackage{multirow}
\usepackage{booktabs}
\usepackage{mathptmx} 
\usepackage{amsmath}
\usepackage{mathtools}
\usepackage{algorithmic}
\usepackage[linesnumbered,ruled,vlined]{algorithm2e}
\usepackage{threeparttable}
\usepackage{multicol}
\usepackage{makecell}
\usepackage{hyperref}
\usepackage{textcomp}
\usepackage{pifont}
\DeclareMathOperator*{\argmax}{arg\,max}

\usepackage[dvipsnames]{xcolor}
\usepackage{pifont}
\usepackage{subfigure}
\newcommand{\ns}[1]{\textcolor{red}{[ns: #1]}}
\newcommand{\mv}[1]{{\textcolor{green}{#1}}}
\newcommand{\wm}[1]{{\textcolor{cyan}{[wm: #1]}}}
\newcommand{\sz}[1]{{\textcolor{orange}{[sz: #1]}}}
\newcommand{\fc}[1]{\textcolor{black}{#1}}

\usepackage{soul}
\usepackage{xcolor}

\newcommand{\rev}[1]{\textcolor{black}{#1}}

\soulregister{\mv}{1}
\soulregister{\ns}{1}
\soulregister{\wm}{1}
\soulregister{\sz}{1}

\begin{document}
%
\title{AIA: A Customized Multi-core RISC-V SoC for Discrete Sampling Workloads in 16 nm}
%
%
%

\author{Shirui~Zhao, ~\IEEEmembership{Graduate Student Member,~IEEE}, Nimish~Shah, Wannes~Meert,~\IEEEmembership{Member,~IEEE},
         and~Marian~Verhelst,~\IEEEmembership{Fellow,~IEEE}%
\IEEEcompsocitemizethanks{\IEEEcompsocthanksitem S. Zhao and M. Verhelst are with the Department
  of Electrical Engineering - MICAS, KU Leuven, Belgium.\protect\\
  \IEEEcompsocthanksitem N. Shah was with the Department of Electrical Engineering - MICAS, KU Leuven, Belgium.\protect\\
  \IEEEcompsocthanksitem W. Meert is with the Department
  of Computer Science - DTAI, KU Leuven, Belgium.}
}

\maketitle



\begin{abstract}
Probabilistic models (PMs) are essential in advancing machine learning capabilities, particularly in safety-critical applications involving reasoning and decision-making. Among the methods employed for inference in these models, sampling-based Markov Chain Monte Carlo (MCMC) techniques are widely used. However, MCMC methods come with significant computational costs and are inherently challenging to parallelize, resulting in inefficient execution on conventional CPU/GPU platforms.
To overcome these challenges, this paper presents AIA, an 
multi-core RISC-V System-on-Chip (SoC) design
fabricated using Intel’s 16~nm process technology. \rev{Our Approximate Inference Accelerator (AIA)} is specifically designed to empower edge devices with robust decision-making and reasoning abilities. The AIA architecture incorporates a RISC-V host processor to manage chip-to-chip data communication and a 2D mesh of 16 custom versatile RISC-V cores optimized for high-efficiency approximate inference.
Each core features (i) custom instructions and datapath blocks for non-normalized Knuth-Yao (KY) 
sampling, as well as for the interpolation of non-linear functions (e.g., logarithmic, exponential), and (ii) direct data-access to the register file of each neighboring core, to reduce the data movement costs of frequent data exchanges between nearby cores. To further capitalize on the parallelism potential in MCMC algorithms, we developed a specialized compile chain that enables efficient spatial mapping and scheduling across the cores.
As a result, AIA attains a peak sampling rate of 1277 MSamples/s at 0.9V and achieves an energy efficiency of 20 GSamples/s/W at 0.7V, surpassing the previous state-of-the-art ASIC accelerator for probabilistic inference by up to 6$\times$ in speed and 5$\times$ in energy efficiency. 
Furthermore, AIA's versatility is demonstrated through the successful mapping of different types of PM workloads onto the chip.
\end{abstract}


%

\section{Introduction}
%
%
%
%

\IEEEPARstart{A}{s} human beings, we have the perception and reasoning capabilities to think fast and slow, to decide based on intuition and to reason \cite{daniel2017thinking}. To empower machines with similar capabilities, emerging computing paradigms such as neural-symbolic AI (NeSy) \cite{nesy} and Bayesian deep learning (BDL) \cite{bdl} have been developed. These methods aim to combine the strengths of neural network based perception, with symbolic or probabilistic reasoning, thereby achieving both data-driven pattern matching and interpretable knowledge-reasoning capabilities.

In recent years, deep neural networks (DNNs) have demonstrated remarkable success in perception tasks, primarily due to their ability to learn from massive amounts of data and their proficiency in handling unstructured data \cite{aggarwal2018neural}. However, they are proven to be ill-suited for reasoning tasks \cite{10682967} \cite{nesy}. In contrast, probabilistic models (PMs) \cite{agrum, pyro, turing, gen.jl} offer robust and interoperable solutions for a wide array of complex decision-making tasks. These models can be learned from less data through the integration of human knowledge and allow people to make decisions taking into account uncertainty.
For instance, Nafar et al. \cite{nafar2024probabilistic} and Feng et al. \cite{feng2024bird} utilize advanced large language models like GPT-4 to extract and understand complex patterns from raw text data sources, subsequently employing Bayesian networks (Bayes Nets) for probabilistic reasoning. This approach has demonstrated critically improved performance compared to sole LLMs. Similarly, 3DP3 \cite{3dp3} enhances 3D scene comprehension for robotic applications by combining the DNN with PM taking into account knowledge of object size and relational dynamics.

However, integrating such machine learning capabilities in embedded systems comes with significant computational challenges. To meet the real-time and low power constraints for edge devices, there has been a growing trend to integrate domain-specific accelerators into the system to overcome the computing and memory burden \cite{pgma, ueyoshi2022diana}. While current hardware developments mainly focus on DNN acceleration, there is a lack of studies exploring the inference acceleration for PM. This paper aims to address this gap by proposing a specialized accelerator for PM inference, which is essential for the advancement of machine learning capabilities in safety-critical applications.

Markov Chain Monte Carlo (MCMC) methods \cite{turing} play a vital role in the probabilistic (or Bayesian) inference for PMs, especially since exact inference is intractable and often simply infeasible for complex models. 
The core of MCMC is to generate samples from a complex target distribution by constructing a Markov chain that converges to it through iterative random transitions, which poses several challenges on off-the-shelf hardware. For example, random number generation (RNG) is not natively supported on most existing hardware platforms, yet it is required to create a large number of uniform random numbers for PM inference. 
Moreover, the sampling operations typically require many random memory accesses, which complicate their parallelization and clutter the memory bandwidth.
Beyond the sampling operation itself, also the wide variety in graph (model) structures of the PMs brings several challenges.
Unlike DNN workloads being completely dominated by regular, general matrix multiplications, different types of PMs can exhibit very different topologies, which can both be \emph{regular} or \emph{irregular}, \emph{directed} or \emph{undirected}. For example, Markov Random Field (MRF) applications \cite{mcmc_cpu}, characterized by regular undirected 2D grid graphs, offer structured memory access patterns and enable massive parallelism during the inference stage, which allows GPUs to significantly outperform CPUs \cite{mrf_uq} for this workload. On the other hand, PMs such as the Gaussian Mixture Model or Bayes Nets \cite{lp_mcmc} are represented by an irregular directed graph and miss any regularity in parallelization opportunities. 
This results in inefficiencies from irregular memory access patterns, which lead to poor cache utilization, suboptimal memory bandwidth usage and hinder memory coalescing. Additionally, parallelizing different parts of the graph across multiple units, whether CPU cores or GPU streaming multiprocessors, incurs significant communication and synchronization overheads.
These unique challenges underscore the necessity for dedicated, yet flexible hardware architectures optimized for probabilistic inference.

State-of-the-art (SotA) DNN accelerators \cite{ueyoshi2022diana,tpu}, unfortunately, are unsuitable for PMs, as they also lack support for irregular computations and specialized randomness operations. Meanwhile, existing probabilistic inference ASIC accelerators only support a narrow subset of PMs \cite{mrf_uq, pgma, dpu}, {due to their fixed function blocks and limited programmability}. This highlights the need for {a new type of processor,} 
capable of general inference across a diverse range of PM types, where flexibility and programmability are balanced with efficiency on both regular and irregular compute patterns.

This paper proposes AIA, a specialized multi-core programmable approximate 
inference accelerator {for PMs}. 
AIA boosts the efficiency of MCMC for various types of probabilistic graphical models, including MRF and Bayes Nets. It was designed to efficiently execute both regular and irregular computational graphs, {and is highly programmable 
to} provide an optimized solution for these emerging workloads. AIA is taped out in Intel 16 nm technology. Its key features are: 
\begin{itemize}
    \item To flexibly support a wide variety of PMs, our design is developed based on programmable, yet customized RISC-V cores with an extended Instruction Set Architecture (ISA), enhanced with custom KY sampling and interpolation instructions built into the execution stage.
    \item To maximize parallelism for throughput, 4$\times$4 parallel accelerator cores are organized in a mesh topology. All cores can asynchronously update different random variables to {simplify the design of the AIA} compiler.
    \item Cores are equipped with access to the register file of neighboring cores for low-cost data synchronization, as frequently required during inference.
\end{itemize}

The paper is organized as follows. The basics of the PM inference algorithm and the related challenges are discussed in {Section} \ref{sec:background}. {Section} \ref{sec:architecture} describes the AIA architecture, the internal details of the customized processor cores, the novel KY sampler, and the hardware interpolation unit. {Section} \ref{sec:compiler} discusses the customized compiler for our hardware. Subsequently, Section \ref{sec:experiments} presents the experimental results and measurements of the taped-out chip and Section \ref{sec:conclusion} concludes the paper.

\section{Background and Challenges} \label{sec:background}
In this section, we briefly introduce the concept of PMs \cite{Koller2009PGMPT}, represented by directed or undirected graphs, {together with} their probabilistic inference through the parallel MCMC sampling, and the opportunities as well as challenges to enable the hardware acceleration for any PM type \cite{andrieu2003introduction}. In this paper, uppercase letters are used to denote Random Variables (RVs), while lowercase letters represent their specific values.

\subsection{Probabilistic Models}
To show the generality of our work, we will outline and discuss two representative PM workloads with widely diverging characteristics: 1.) 
Bayes Nets, represented by directed irregular graphs. 
and 2.) 
MRFs, represented by undirected regular 2-D grid graphs. 

\begin{figure}[!t] 
\centering
\includegraphics[trim={0cm 0cm 0cm 0cm} , clip, width=0.75\columnwidth]{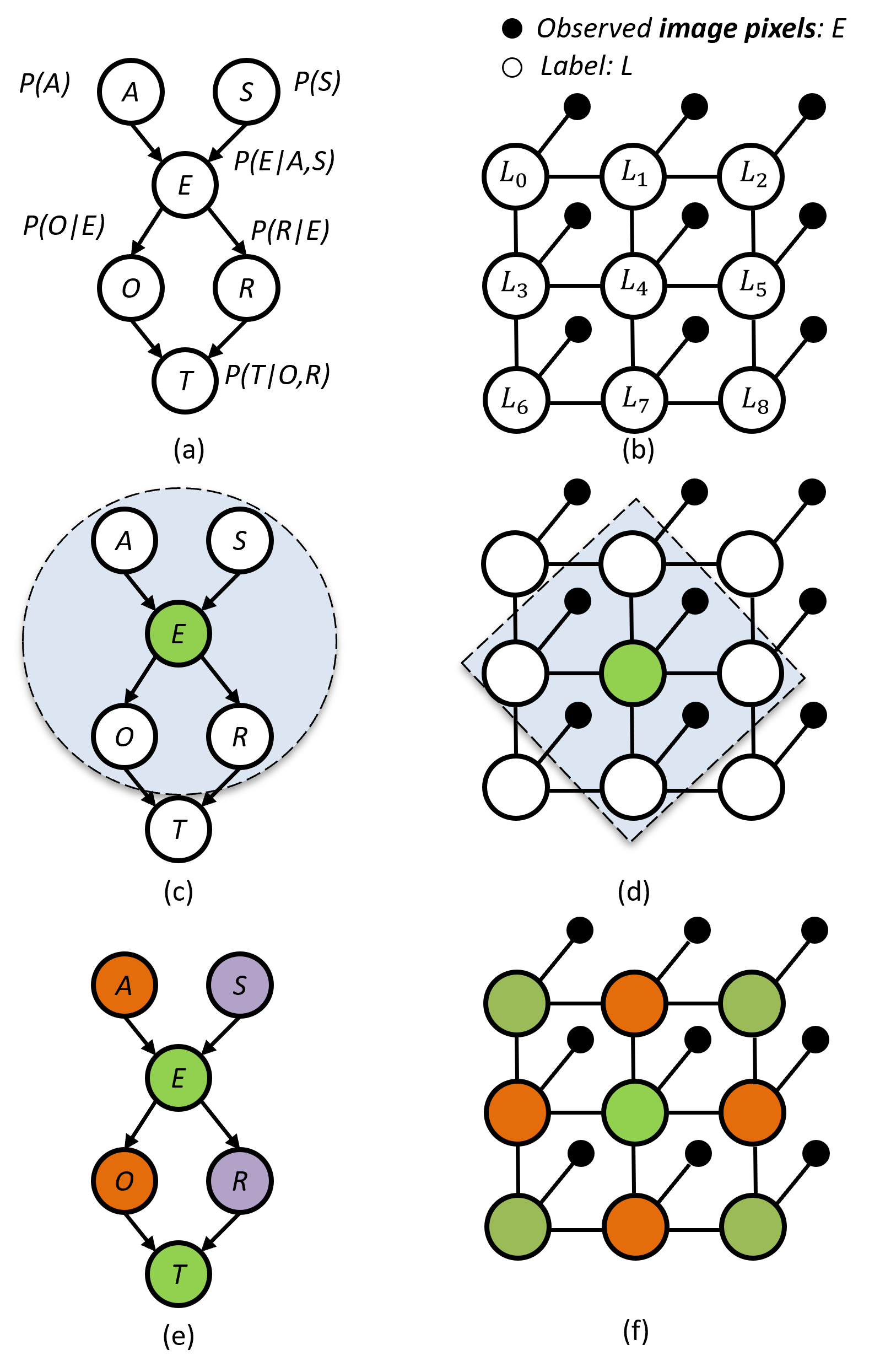}
\caption{This figure provides an illustrative breakdown of RV-level parallelization. (a-b) \rev{Example of a directed graph (Bayes Net) and an undirected graph (MRF)}. (c-d) The Markov blanket (MB) of green RV. Given its MB, the node is conditional independent with all the other nodes. (e-f) Following the MB property, the conditional independent RVs (nodes with the same color) can be updated simultaneously.}
\label{fig:rv_parallel}
\end{figure}

\subsubsection{Directed graphs}
The Bayes Net, as shown in Fig.~\ref{fig:rv_parallel}(a), is represented as an irregular directed acyclic graph (DAG), whose nodes are RVs and directed edges between nodes represent the mutual RV relationships, in the form of conditional dependen
An edge from node \textit{A} to node \textit{B} indicates that \textit{A} directly influences \textit{B}. For this, each node is associated with a conditional probabilistic table (CPT) that quantifies the effect of the node's parents on the RV. Using these CPTs, the joint probability distribution function $P(X_1, X_2, ..., X_n)$ for a set of RVs $\{X_1, X_2, ..., X_n\}$ can be factored as:
\begin{equation}
    P(X_1, X_2, ..., X_n) = \prod_{i=1}^{n} P(X_i|\pi(X_i))\fc{,}
\end{equation}
where $\pi(X_i)$ are all the parent RVs of \textit{$X_i$}, and $P(X_i|\pi(X_i))$ is the associated CPT. 

Given a Bayes Net, a query refers to the process of computing probabilities of one or more variables of interest, given the information about other variables. Two kinds of queries exist for the Bayes Nets:

\textbf{Marginal probability queries} aims at computing the probability of a single variable or a set of variables. For example, in a medical diagnosis Bayes Net that models diseases and laboratory tests, $P(D,T)=P(T|D)P(D)$, $D$ represents RV $Disease$ which has two values: $True$ or $False$, and $T$ is the test result which could be $Positive$ or $Negative$. To simplify the notation, $d$ and $t$ represent the patient who has the disease and the test is positive, respectively, while $\overline{d}$ and $\overline{t}$ denote no disease and a negative test. 
To compute the probability that any can have a 
particular disease, say $P(d)$, it requires summing out all possible values of the other variables in the network that are related to the disease variable, or in this case: $P(d)=\sum_{T} P(d, T) = P(d,t) + P(d,\overline{t})$. 

\textbf{Conditional probability queries} inquire about the probability of a variable or a set of variables given some evidence. Following the previous example, given that a patient tests positive (evidence), the query could be deriving the probability of the disease being true. Mathematically, it is expressed as $P(d|t)$, which can be computed via Bayes theorem:
\begin{equation}
    P(d|t) = \frac{P(d,t)} {P(t)} = \frac{P(d)P(t|d)} {P(d)P(t|d)+P(\overline{d})P(t|\overline{d})}.
\end{equation}




\subsubsection{Undirected graphs}
In an undirected graph of a probabilistic model, such as a Markov network \cite{van2012markov}, Boltzmann machine \cite{fischer2012introduction}, the nodes again represent a set of RVs, and the edges indicate the probabilistic dependencies or constraints between the connected RVs. Yet, here these dependencies are not directional. A MRF is a type of such undirected PM, as shown in Fig. \ref{fig:rv_parallel}(b).


A MRF defines the joint probability of variables $X=\{X_1, X_2,...,X_n\}$ \cite{coopmc, van2012markov}, which is represented as:

\begin{equation}\label{eq:mrf_join}
P(X=x) = \frac{\prod_{c \in Cliques}\phi_{c}(x_c)}{Z}=\frac{exp({\sum_{c}f_{c}(x_c)})}{Z}\fc{,}
\end{equation}
where {$x$ is the joint state of the variables $X$,} $x_c$ is the state of the clique $c$ as defined by the graph, $\phi_c$ is a \textbf{potential function} over the RVs in clique $c$, and $Z$ is the normalization constant. {Due to its large dynamic range, this joint probability} is often computed in the $logarithmic$ domain, where the clique potential function is replaced by the exponential sum of an \textbf{energy function} $f_c$ \cite{mrf_uq, pgma,parallel_gibbs} {(see further at Eqn. (7) for a specific example)}.

The main inference task in an MRF is to compute the Maximum Posterior Estimation (MPE), which requires finding the most probable assignment of all unobserved variables given evidence. For example, in image denoising task \cite{coopmc} (shown in Fig.~\ref{fig:rv_parallel}(b)), the RV $X$ can be divided into two parts: a clean image (with unobserved label variables $L$ represented with \textbf{hollow circle}) associated with the noisy input image (with observed variables $E$ carrying the value of the evidence $e$ represented with \textbf{solid circle}). The objective of MPE is to find the most likely {state $l$} 
of the entire clean image:
\begin{equation}\label{eq:mrf_mpe}
l^*=\argmax_{l}P(L=l, E=e)\fc{.}
\end{equation}

\begin{figure*}[!t]
    \centering
    \includegraphics[trim={0cm 0cm 0cm 0cm} , clip, width=\textwidth]{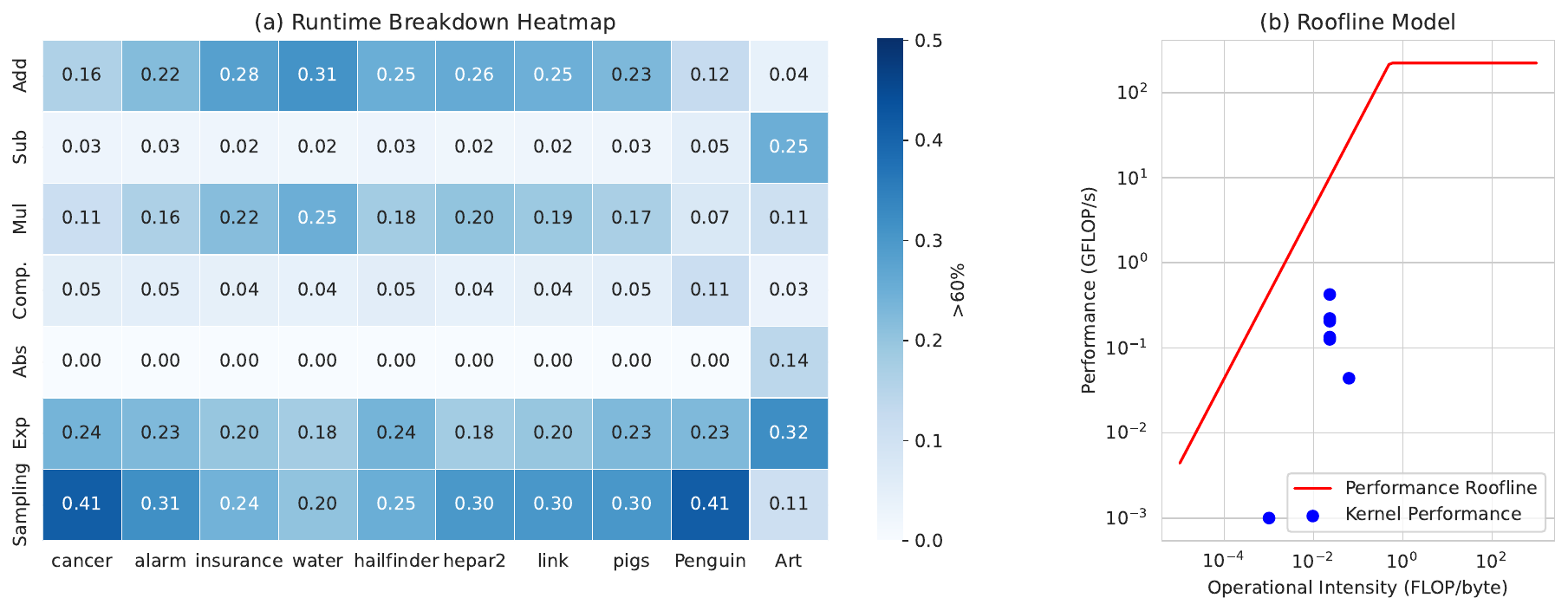}
      \caption{\rev{(a) Runtime characterization of different benchmarking workloads, highlighting the diverse operations necessary for probability distribution computation, with sampling identified as the critical operation. (b) Roofline model analysis conducted on an Intel i7-7800X CPU @ 3.50GHz, demonstrating that Gibbs sampling is memory-bound. This suggests the need for a custom accelerator with improved data transfer and fused operation features for these applications.} }
    \label{fig:op_breakdown}
\end{figure*}

\subsection{Probabilistic inference} 
To evaluate marginal, conditional, or MPE queries on the PMs, 
Bayes' theorem has to be applied across all the involved conditional relationships of the graph. 
There are multiple popular methods to perform this inference, depending on the complexity of the PM \cite{pml_nature}. For tractable models, exact inference algorithms, such as \textit{variable elimination} \cite{sanner2012symbolic}, \textit{weighted model counting} \cite{dice}, \textit{message passing} \cite{jordan2002graphical}, can be used. However, these exact inference methods can be computationally intractable or impractical for many real-world problems \cite{jordan2002graphical}. The computation often has a time complexity that scales poorly with the data size or the number of variables in the model, and evaluating the normalization constant $Z$ could also be computationally infeasible \cite{pml_nature}. 

In this case, approximate techniques such as Markov Chain Monte Carlo (MCMC) methods bring the remaining feasible solution for flexible models. MCMC algorithms use samples to approximate complex distributions. In this work, we focus on Gibbs sampling \cite{parallel_gibbs}, as one widely used MCMC algorithm to infer PMs.

\label{sec:parallel_inf}
\begin{algorithm}\label{alg:seq_gibbs}
    \caption{Sequential Gibbs Sampling}
    \For{$Chain$ $chain = 1$ to $MaxChain$}{
        \For{$Iteration$ $t = 1$ to $T$}{
        \ForAll{$Variables$ $X_i \in X$}{Execute Gibbs Update:\\ $x_i^{t} \sim 
        P(x_i^{t} | x_1^{t},..., x_{i-1}^{t},x_{i+1}^{t-1},...,x_n^{t-1})$} 
        }
    }
\end{algorithm}

\begin{algorithm}
\caption{Parallel Gibbs Sampling}\label{alg:parall_gibbs}
\begin{algorithmic}
    \FOR{$Chain\ chain = 1$ to $MaxChain$}
        \FOR{$Iteration\ t = 1$ to $T$}
            \FOR{$Color\ k = 1$ to $K$}
                \STATE \textbf{parallel for} $Variable\ X_i \in X_c$ \textbf{do}
                \STATE \hspace{1em} Execute Gibbs Update:
                \STATE \hspace{2em} $x_i^{t} \sim 
                P(x_i^{t} | x_1^{t}, \dots, x_{i-1}^{t}, x_{i+1}^{t-1}, \dots, x_n^{t-1})$
                \STATE \textbf{end parallel for}
            \ENDFOR
        \ENDFOR
    \ENDFOR
\end{algorithmic}
\end{algorithm}

\subsubsection{Sequential Gibbs sampling}
Sequential Gibbs sampling is built out of the sampling chain, in which in 
each chain, variables are sampled one at a time in a fixed sequence. Each Gibbs update step involves the updating of one variable, given the state of all other variables. When all variables $X$ are updated in this way, one iteration is done. Sampling has to continue for a sufficient number of such iterations $N$ to ensure the chain has converged or "mixed" (i.e., reached its stationary distribution $P(X)$). Optionally, multiple such chains could run in parallel, as summarized in Alg. \ref{alg:seq_gibbs}. 
 
 The Gibbs update for each variable \(X_i\) involves two steps: \textcircled{1} computing the conditional distribution \(P(x_i \mid x_1, \ldots, x_{i-1}, x_{i+1}, \ldots, x_n)\), and \textcircled{2} sampling from this distribution (denoted by \(\sim\)) to assign the result to the current variable \(X_i\). 
\rev{For our design, we used an ALU and a custom interpolation unit for the probability distribution computing. Sampling operations are processed by a hardware sampling unit.}

 The computation in step \textcircled{1} can be simplified by leveraging the Markov Blanket (MB) \cite{mb} properties of the graph. The MB of a node in PMs is the set of nodes that "shield" it from the rest of the network, shown in Fig. \ref{fig:rv_parallel}(c-d). 
 According to the conditional independence rule, the Gibbs Update step in Alg.~\ref{alg:seq_gibbs} (and later also for Alg.~\ref{alg:parall_gibbs}) can be simplified:
\begin{equation} \label{eq:mb}
    P(X_i|X_{\backslash i})= P(X_i|MB(X_i)),
\end{equation}
where $X_{\backslash i}$ represents all the other RVs except $X_i$, ${MB}(X_i)$ denotes the MB of $X_i$ in the graph. This reduces the computational cost since it only depends on the value of $X_i$'s MB, instead of the entire graph.

\textbf{Directed graph}.
In the directed graph, the MB of a node consists of its parents, children, and children's parents. 
Following the Eqn.~\ref{eq:mb}, the corresponding conditional probability of the directed graph is defined as:
\begin{equation}\label{eq:bn}
    P(X_i|MB(X_i)) \propto
P(X_i|\pi({X_i}))\prod_{C \in Children(X_i)}{P(C|\pi({C}))}\fc{.}
\end{equation}
in the example of Fig.~\ref{fig:rv_parallel}(a), let's assume the values of RVs are $\{a, s, e, o, r, t\}$ in a certain step, then the distribution required to update RV $O$ is: $P(O|e,t,r) \propto P(O|e)P(t|O,r)$.

\textbf{Undirected graph}.
The MB of a node in an undirected graph is slightly different, which consists of only its direct neighbors. 
The example model in Fig.~\ref{fig:rv_parallel}(d) could be used for image denoising, the MB of the current label node $L_{i}$ consists of 4 neighbor label nodes $L_j$ and one input pixel node $E_{i}$. 
The conditional probability is given as \cite{pgma}:
\begin{align}\label{eq:mrf}
P(L_i|MB(L_i)) \propto exp(\sum_{j}\theta_{i,j} L_iL_j+h_{i}L_iE_i),
\end{align}
where $\theta_{ij}$ and $h_i$ are the parameters for the smoothening cost (neighboring pixel interactions) and the data cost (label-evidence interaction), 
respectively. $j$ represents the index of the neighbor label nodes, which e.g. interact in the range $[1,4]$ to represent its four neighbors. Given the neighbors and evidence, it requires assigning RV $L_i$ to all possible values to generate the distribution.

After the computation of this distribution, step \textcircled{2} of the Gibbs update involves the $sampling$ operation from this probability distribution with or without normalization according to the sampling algorithm requirement. Traditional sampling algorithms, like cumulative distribution sampling (CDF) \cite{pgma} need to normalize the distribution to ensure they sum to 1. In our KY sampler (see section \ref{sec:se_unit}), the normalization step can be avoided.


\subsubsection{Parallel inference}
Within a single sampling chain, throughput, and hence convergence, can be improved by further parallelizing variable updates. 
For this, we can again rely on the MB property. As random variables that are conditionally independent do not exhibit data dependencies, they can be updated simultaneously rather than sequentially. 
The parallel Gibbs sampling algorithm \cite{parallel_gibbs} exploits this RV-level parallelism, where a group of variables is jointly updated.

Which group of variables could be updated together, is determined by coloring the nodes in such a way that each node is assigned one of \textit{K} colors in which all nodes with dependencies have different colors, as shown in Fig.~\ref{fig:rv_parallel}(e) and (f). 
As all variables of the same color are conditionally independent from each other, they can be computed and sampled simultaneously. As such, for a given color, the parallel Gibbs sampler simultaneously draws new values for all random variables before updating the nodes with other colors, as outlined in Alg.~\ref{alg:parall_gibbs}. 

Regular 2D grid models such as an MRF \cite{pgma, mrf_uq} (shown in Fig.~\ref{fig:rv_parallel}(f)), can be realized with a 2-color parallel sampling flow. While irregular PMs (such as Fig.~\ref{fig:rv_parallel}(e)) might need more colors in their parallelization scheme.


\begin{figure*}[!t]
    \centering
    \includegraphics[trim={0cm 0cm 0cm 0cm} , clip, width=\textwidth]{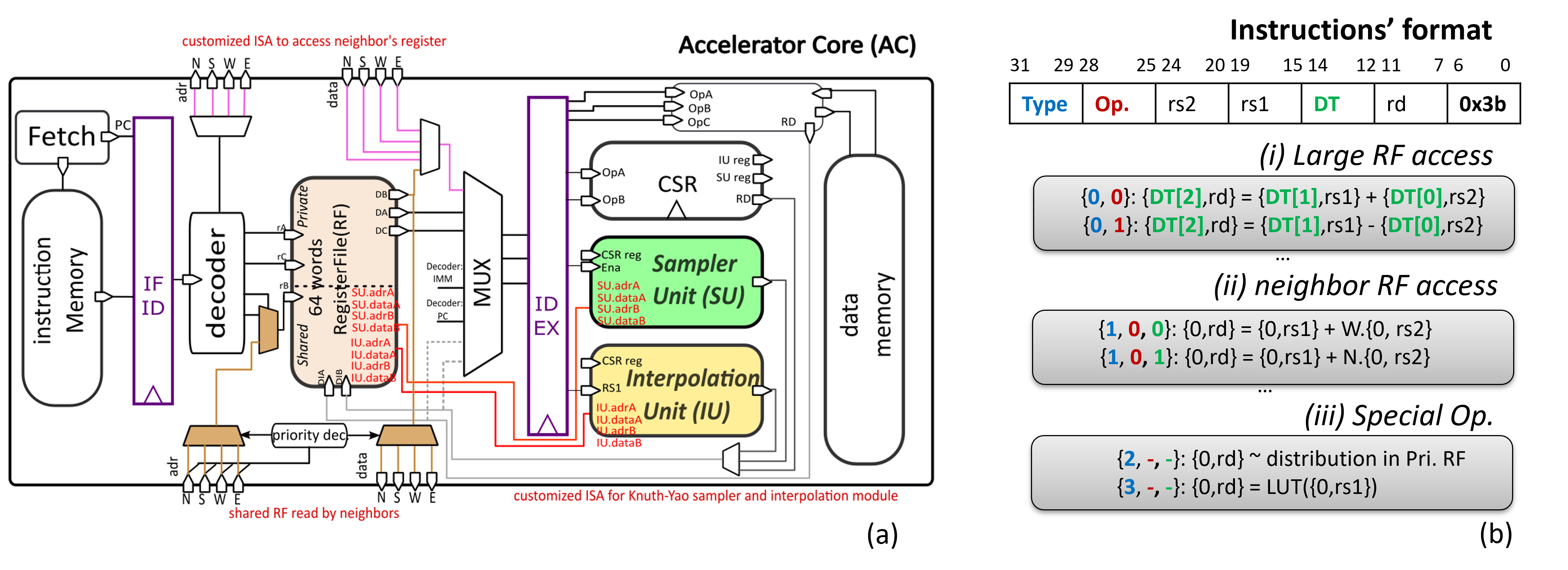}
    \caption{(a) The architecture of accelerator core. The ALU, multiplication, and dot-product unit of the baseline \rev{RI5CY} are also present in the EX stage, but not shown for clarity. (b) Custom ISA for: large RF access, neighbors' shared RF access, and special operations like sampling and lookup table function.}
    \label{fig:ac_arch}
\end{figure*}
\subsection{Challenges to support general PMs}
Supporting general PMs, especially irregular ones, poses several challenges for efficient execution on general-purpose mutli-core processors like CPUs, GPUs and existing probabilistic inference accelerators. To understand the requirement to accelerate the previously mentioned probabilistic models and get insights for hardware design, we profile the representative workloads in terms of runtime with the open-sourced framework, AGrUM \cite{agrum} on Intel i7 CPU. 

\textbf{Runtime breakdown}. Fig.~\ref{fig:op_breakdown}(a) \rev{illustrates the runtime distribution for ten workloads. Key observations include: (1) The algorithms utilize a diverse set of operations, such as Addition (Add), Subtraction (Sub), Multiplication (Mul), Comparison (Comp.), Absolute Value (Abs), Exponentiation (Exp), and Sampling. (2) The most computationally intensive operations vary across different workloads. (3) The Sampling operation is the most time-consuming, accounting for nearly half of the total runtime.}

\textbf{Roofline Model Analysis.}
Fig.~\ref{fig:op_breakdown}(b) \rev{shows the roofline model analysis result of the workloads on an Intel i7-7800X CPU using Intel Advisor 2025} \cite{advisor}. \rev{The analysis indicates that all workloads are memory-bound. This is primarily due to the sampling-based algorithms, which involve a series of element-wise operations along with frequent memory accesses to store and load distributions, leading to low hardware utilization.}

\rev{We further summarize the challenges of the acceleration of approximate inference} in Table \ref{tab:challenges}, along with AIA's related innovations to address them.
\renewcommand{\arraystretch}{1.2}
\begin{table}[!t]
\centering
\caption{Challenges and opportunities in parallel Gibbs sampling and related AIA innovations
}
\begin{tabular}{ll}
\toprule
\multicolumn{1}{c}{\begin{tabular}[c]{@{}c@{}}\textbf{Challenges/opportunities}\\ \textbf{for general PMs}\end{tabular}} & \multicolumn{1}{c}{\textbf{AIA innovations}} \\ \midrule
\begin{tabular}[c]{@{}l@{}}
Diverse applications with \\ varying graph structure \end{tabular} &
\begin{tabular}[c]{@{}l@{}}
Programmable ALU to support rich \\ operations \end{tabular} \\
\hline
Frequent synchronizations &\begin{tabular}[c]{@{}l@{}}Hardware synchronization \\ with event unit, and \\ {core-to-core Register File (RF) access} \end{tabular} \\ 
\hline
SIMD unfriendly &\begin{tabular}[c]{@{}l@{}}Asynchronous accelerator cores \\ with independent instructions\end{tabular} \\  
\hline

Inefficient sampling & hardware KY sampler \\ 
\hline
\begin{tabular}[c]{@{}l@{}}
Nonlinear operations \\ $(exp, log, etc.)$ \end{tabular} & hardware interpolation unit\\ 
\bottomrule
\end{tabular}%
\label{tab:challenges}
\end{table}
 \begin{figure}[!t]
    \centering
    \includegraphics[trim={0cm 0cm 0cm 0cm} , clip, width=0.8\columnwidth]{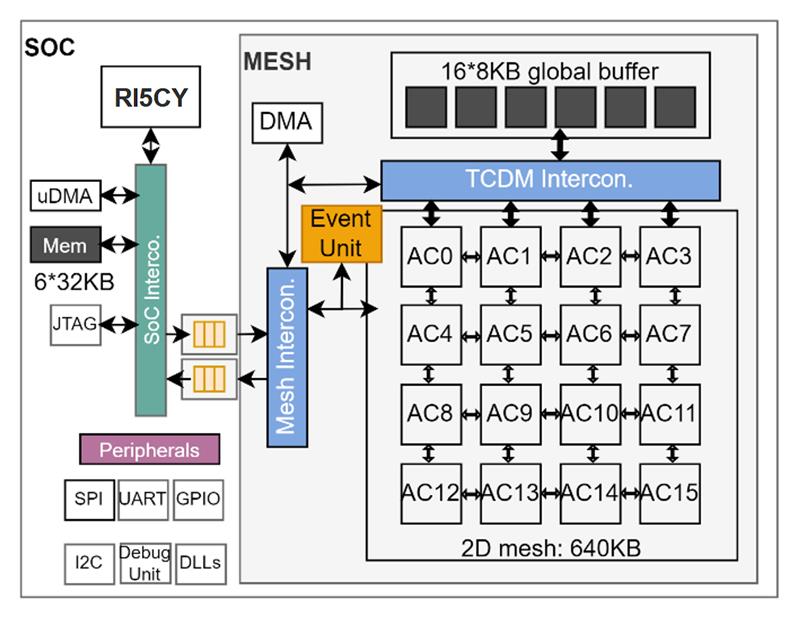}
    \caption{The AIA architecture with 16 accelerator cores.}
    \label{fig:top_arch}
\end{figure}
\subsubsection{\textbf{Poor programmability of existing accelerators}}
The diverse nature of PMs demands flexible hardware solutions that can adapt to different computational patterns and model topologies, making programmability a key requirement for efficient acceleration.
However, existing PM ASIC acceleration only targets regular workloads \cite{pgma, mrf_uq}. FPGA accelerators \cite{acmc2} are flexible but require generating different bitstreams for different models and lack energy efficiency compared to ASIC solutions. AIA's processor, on the other hand, is fully SW-programmable.

\subsubsection{\textbf{Frequent Synchronizations}} \label{sec:frequent_sync}
Parallel MCMC sampling requires frequent updates between nodes to exchange the values of the updated variables, which can introduce costly synchronization overheads for CPUs/GPUs. \rev{By leveraging the core-to-core backdoor Register File (RF) access pattern, data can efficiently be shared between neighboring cores. From the control side, AIA's event unit will serve as a hardware synchronization mechanism, reducing power consumption by disabling core clocks when idle.} 

\subsubsection{\textbf{SIMD unfriendly}} 
The nodes with the same color, which will be executed in parallel, may perform the same arithmetic operation on the inputs, rendering them suitable for a SIMD execution. However, the inputs of these nodes typically reside in random memory locations, owing to the irregularity of the graph structure. 
Experiments in \cite{parallel_gibbs} find that around 50\% of the load requests in graph workloads result in cache misses. This leads to high variability in the load latency of these inputs, causing all the SIMD lanes to stall for the slowest input. 
Thus, despite the availability of parallel operations to execute, random memory loads make irregular graphs inefficient on \rev{general purpose SIMD-oriented architectures, such as} multi-core CPUs and multi-threaded GPUs. AIA \rev{therefore} uses asynchronous accelerator cores that execute independent instructions instead of a SIMD unit.

\subsubsection{\textbf{Inefficient Sampling and Complex Operations}}
 Sampling is a core operation in MCMC-based inference, with frequent use of complex mathematical functions to first compute the sampling probability distribution of the energy function (e.g., exponential in Eqn.~\ref{eq:mrf}). The AIA architecture includes specialized hardware for fast, energy-efficient sampling (via a KY sampler) and an interpolation unit for complex function approximations, reducing computational overhead.

\section{AIA architecture} \label{sec:architecture}
To overcome the previously mentioned challenges, we propose the 16 nm AIA SoC with a \rev{ CV32E40P (RI5CY)} host core \cite{pulp} and a 4$\times$4 mesh of customized accelerator cores (AC) to accelerate sampling workloads. AIA, shown in Fig. \ref{fig:top_arch}, is organized in two clock domains, SOC and MESH domain, which communicate through a cross-clock FIFO.
The host core in the SoC domain is as same as the host of Marsellus \cite{marsellus}. It is responsible for connecting the accelerator to the outside world via peripheral interface and it contains a 192KB SoC memory to buffer data. 

The MESH domain contains 4x4 accelerator cores with mesh connections between them. Each core is a modified RISC-V processor implementing a custom ISA. The new ISA supports additional instructions for sampling, lookup table interpolation, and mesh access. This is enabled by adding a dedicated hardware sampler unit, linear interpolation unit, and an enlarged register file (private/shared RF) to the RISC-V core. The detailed architecture of the cores is discussed in the following section. Each core has its own instruction memory as well as local data memory and the capability to \textit{independently} execute instructions. 
An event unit is implemented for multicore synchronization. The tightly coupled 128KB global buffer is accessible by the top 4 cores only, to reduce the complexity of the Tightly Coupled Data Memory (TCDM) interconnect by $4 \times$ compared to the 16-core 1-D configuration of \cite{marsellus}.

\subsection{Accelerator core architecture and \texttt{Xprob} custom instructions}
The main contributions of AIA on the hardware side are the accelerator core architecture together with its supporting ISA instruction extensions, named \texttt{Xprob}.
Each accelerator core, shown in Fig. \ref{fig:ac_arch}(a), is developed based on the open source \rev{ RI5CY} design \cite{pulp}. Yet, it is enhanced with custom accelerators in the execution stage to accelerate the inference steps. To accommodate the generated intermediate probability distributions and required special function lookup tables, the RF size is increased from 32 words to 64 words. Moreover, data can be fed into the execution stage no only from the register file of the core itself but also from the lower 32-word section of the enlarged register file of the neighboring cores. 
To support and control these new hardware features, 
\texttt{Xprob} ISA extensions add new instructions to the \texttt{XpulpNN} instruction set \cite{marsellus}. Here, the op-code 0$\times$3b is used to identify the extended instructions, while instruction bits [31:29] indicate the new instruction type, as shown in Fig. \ref{fig:ac_arch}(b).
\begin{figure*}[!t]
\centering
\includegraphics[trim={0cm 0cm 0cm 0cm} , clip, width=0.95\textwidth]{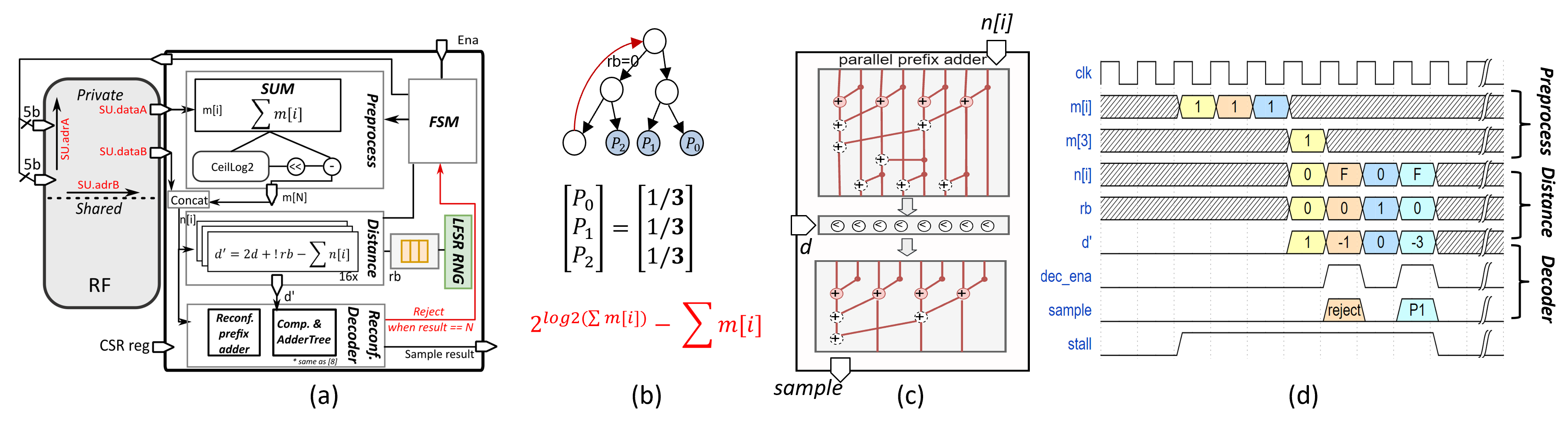}
\caption{\rev{(a) The proposed rejection-based KY sampler hardware architecture. (b) DDG tree for uniform distribution: $P_i=1/3$ and the formal to compute the rejection item as shown in the preprocess submodule. (c) Reconfigurable decoder architecture for different precision of $n[i]$. (d) The signal waveform when random bits from LFSR is $0010$. If the first sample result is a rejection, then the FSM will control the hardware to re-sample.}}
\label{fig:se}
\end{figure*}%

\rev{We modified the ISA to let all R-type (Register-type) RISC-V ALU operations, including addition, subtraction, xor, or, and, shift operations, support access to both the local and shared register files of their neighbors. This is indicated by the instructions. Hence} \textcolor{blue}{\textbf{Type}}\rev{-field, in which a} \textcolor{blue}{\textbf{Type-0}} \rev{points to the local RF with its own private and shared RF, while} \textcolor{blue}{\textbf{Type-1}} \rev{ISA are used to access its neighbors' shared RF.  All Type-0 and Type-1 R-type ALU operations involve three register operands: two source registers ($rs1$, $rs2$) and one destination register ($rd$).}

\textbf{Access Private RF.}
In \textcolor{blue}{\textbf{Type-0}} instructions, the three bits of the \textcolor{Green}{\textbf{DT}}-field here indicate whether the data of the $rs1$, $rs2$ and $rd$ each respectively come from the private or the shared section of the local RF. 
\rev{An example instruction to write the add two local shared register together and write result to private RF could be:}
\textit{Xprob.add.hll x1, x1, x2}.
\rev{The "hll" suffix means that the destination register ($rd$) is in private RF, while the source registers ($rs1$, $rs2$) are in shared RF. Similarly, different suffixes like "hhl," "lhh," etc., adjust the DT field appropriately to indicate the registers are in shared or private RF.} 

\begin{figure}[!t]
    \centering
    \includegraphics[trim={0cm 0cm 0cm 0cm} , clip, width=\columnwidth]{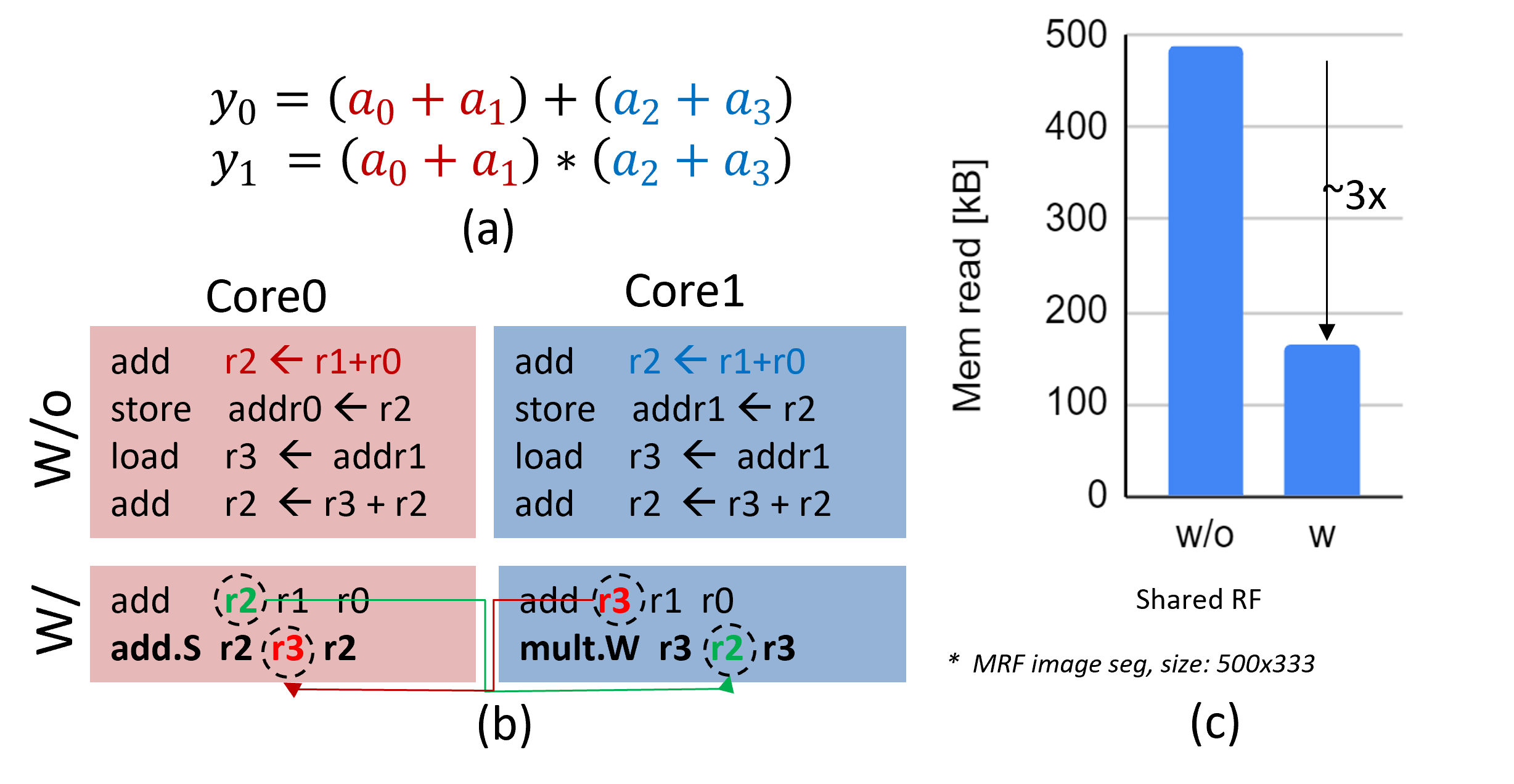}
    \caption{\rev{(a) A toy example to show how to access neighbors' shared RF. (b) Pseudo-assembly code for the ISA without (W/o) and with (W/) neighboring RF access. (c) With this feature, the memory access pattern can be reduced by $3\times$ for a real-world MRF application.}}
    \label{fig:shared_ISA}
\end{figure}

\textbf{Access Neighbors' shared RF.}
\textcolor{blue}{\textbf{Type-1}} instructions are introduced to enable access to the shared part of the register file of neighboring cores to fetch one of the operands ($rs2$) of an R-type instruction. 
Here, \textcolor{Green}{\textbf{DT}} signals whether this data should be fetched from West (\textcolor{Green}{\textbf{DT}=0}), North (\textcolor{Green}{\textbf{DT}=1}), South (\textcolor{Green}{\textbf{DT}=2}), or East (\textcolor{Green}{\textbf{DT}=3}) core's shared RF, respectively. During the data sharing, all the cores with access to their neighbors shared registers simultaneously. \rev{Fig.}~\ref{fig:shared_ISA} \rev{illustrates with a toy example this feature of accessing the neighbors' shared RF}. For the example workload shown in Fig.~\ref{fig:shared_ISA}(a), \rev{we can avoid to exchange the data between cores through the shared global memory, as would be required in the baseline multi-core CPU design. Instead, the compiler can introduce a} \textcolor{Blue}{\textbf{Type-1}} \textit{add.S} instruction 
with \textcolor{Green}{\textbf{DT}} \rev{as $2$ to directly read the $West$ neighbor's shared RF. With this feature, two load/store to global buffer operations are eliminated and 3$\times$ memory read pattern can be reduced for real-word MRF application, as shown in Fig.~}\ref{fig:shared_ISA}(b-c).

\textbf{Special Operations.}
The remaining instruction \textcolor{Blue}{\textbf{Type-2}} and \textcolor{Blue}{\textbf{Type-3}} support the specialized instructions for the Sampling Unit and Interpolation Unit in the execution pipeline stage. The multi-cycle \rev{sampling} instruction will halt the RISC-V pipeline until the sampling operation is finished and then write the sample result to $rd$. The interpolation instruction will finish the lookup operation in just one cycle.

\textbf{Update RISC-V GCC.}
\rev{We modified the binutils in PULP RISC-V GNU Compiler Toolchain }\cite{pulp_gcc} \rev{to recognize these instructions and rebuild the toolchain to support them. In the practical implementation, inline assembly code is used for the custom instructions.}

\textbf{Approximate Inference Overview.}
During PM inference, each AIA core iteratively works as follows: 
\begin{itemize}
    \item \rev{Exchange information in terms of RV values between different cores via the neighbor RF access ISA. All cores will maximally take four cycles to read its four neighbors' updated labels in their shared RF.}
    \item \rev{Utilize the ALU through the standard RISC-V ISA for the norm computation.}
    \item \rev{Use the interpolation unit for nonlinear function computation to compute the smoothness and/or data costs for the probability distributions.}
    \item \rev{Store the resulting probability distribution parameters in the private RF using the large RF access instructions.}
    \item  \rev{Trigger the sampling unit to perform the sampling of the random variable Gibbs update step and write the result into shared RF.}
\end{itemize}
We will now dive into more detail for each of the newly introduced hardware blocks.

\subsection{Enlarged RF}

MCMC frequently requires access to its neighbor RVs, hence benefiting from a 2D mesh architecture for data communication between processing cores dealing with different nodes in parallel. 
We enlarged the register file (RF) size from 32 words to 64 words and divided it into two sections: a private RF and a shared RF. Additionally, an extra read and write port is added for the sampler unit, and another read port is added for the interpolation unit.

\textbf{Private RF}: The regular R-type instructions can only access data in its own shared RF. The \textcolor{blue}{\textbf{Type-0}} \rev{instructions in the ISA are designed to access the private RF. This RF section is used for storing the intermediate results during the distribution computing in the Gibbs update step.} As the distribution will immediately be consumed in the next sampling step, we store the resulting probabilities in private RF to reduce the memory footprint. We also store the 32-word LUT entries in private RF for ithe nterpolation unit.

\textbf{Shared RF}: Each accelerator core contains a shared register file with a size of 32 words. The \textcolor{blue}{\textbf{Type-1}} \rev{instructions in the ISA are designed to access the neighboring shared RFs. After the sampling operation, the sample result will be stored in the shared RF for the neighboring cores to read.} This part of the register file is accessible by the neighboring $N,E,S,W$ cores. This is enabled through an additional register file read port. Only one core at a time can read from this register file section. The priority encoder is used for the multiplexer of read addresses from neighbors and returns data to neighbors. \rev{The custom compiler is carefully designed to avoid colliding accesses and ensure at compile time that there are no conflicts. During the information exchange stage, all the cores will simultaneously access their $N,W,S,E$ neighbor's shared RF. By doing so, we keep the priority decoder with a one-hot selection input and as such avoid collisions.}

\subsection{Sampler Unit} \label{sec:se_unit}
\rev{As described in Sec.~}\ref{sec:background}\rev{, we require sampling operations after the distribution generation. In this section, we show the hardware design of the hardware sampling unit.} The sampler unit for sampling from the distribution is developed based on an optimized hardware-friendly KY algorithm \cite{fldr}.

The initial Knuth-Yao (KY) algorithm transforms a discrete sampling problem into a series of bit operations and table lookups. It is particularly useful for distributions with a finite number of outcomes and is widely used in cryptographic applications due to its efficiency\cite{ky_sampling, ky_sampling_fpga}. \rev{An efficient implementation of such KY hardware sampler based on normalized distributions can be found in }\cite{sz_date}. Here, we propose \rev{an improvement upon the design of} \cite{sz_date} \rev{expanding it with} rejection-based sampling (Fig. \ref{fig:se}) to efficiently sample from discrete distributions without requiring the costly distribution normalization step.

\textbf{Preprocess.}
Our rejection-based KY sampling algorithm starts from a discrete probability distribution $\{P_0, P_1, ..., P_n\}$ as $\{m_0, m_1, ..., m_n\}$, where $P_i=m_i/\sum_i{m_i}$. 
This can be transformed into an equivalent distribution vector consisting of the elements $m_i$, all integer numbers. 
For example, the uniform distribution $P_i=1/3$ with 3 bins in Fig. \ref{fig:se}(b) can be represented as: $m_i=1$. The rejection \rev{probability} $rej/2^w$ can be determined based on the precision $w$ of the distribution, with:
    \begin{align}
        w &= \lceil log2(\sum_{i}{m_i}) \rceil, \\
        rej/2^w &= 1 - \sum_{i}{m_i/2^w},
    \end{align}
$rej/2^w$ \rev{indicates which fraction} of the samples will have to be discarded and retried. 
The updated distribution can now be represented as $\{m_0, m_1, ..., m_n, \textcolor{red}{rej} \}$. In the example of Fig. \ref{fig:se}(b), the precision $w$ is $2$ and the rejection probability is $1/4$.

\rev{To compute $rej$, the sampling unit reads the probability matrix in a row-wise fashion via the RF port $SU.A$ shown in Fig.} \ref{fig:se}(a). 

\textbf{Distance computing.}
This updated distribution has the benefit that all its probabilities sum up to a power of 2. This allows us to make use of a discrete distribution-generating (DDG) decision tree~\cite{fldr}. The sampling process now consists of traversing this tree with random bits. Shown in the example of Fig.~\ref{fig:se}(a), it starts at the root of the tree. If the first random bit is $0$, move left; otherwise, move right. Once a leaf node is reached, the corresponding label is the sampled value. If a rejection node is hit, the sampler rejects the sample, returns to the root node, and performs the sampling again. 
The number of operations and random bits required to sample a value is related to the depth of the tree, which is usually much smaller than the number of possible values in the distribution.

To execute this sampling operation in hardware, we transform the decision tree into a binary matrix in which each element $s_{ij}$ represents the binary value of the distribution element $m_i$ at level $j$ of the tree. Hence the example in Fig. \ref{fig:se}(a) has a binary matrix $M$:

    \begin{equation} \label{eq:prob_matrix}
        M_{4\times2}=
          \begin{bmatrix}
             \textcolor{red}{0} & \textcolor{red}{1}\\
             0 & 1 \\
             0 & 1\\
            0 & 1 \\
          \end{bmatrix}.
    \end{equation}


\begin{table}[!t]
\centering
\caption{\rev{Performance comparison with an state-of-the-art sampler unit}}
\begin{tabular}{lcccc}
\toprule
 & \multicolumn{3}{c}{This work} & CDF \cite{lp_mcmc} \\ \midrule
Operating mode & 32b & 16b & 8b &  32b \\ 
Area ($\times 10^3$ $\mu$m$^2$) & \multicolumn{3}{c}{5.2} & 8.4 \\
Energy (pJ/sample) & 62.5 & 21.7 & 14.4 & 191.3  \\
Throughput (sample/cycle) & 1 & 2 & 4 & 1  \\
\bottomrule
\end{tabular}%
\label{tab:ky_cdf}
\end{table}

This matrix is stored in the private register file and used in the sampler hardware, whose micro-architecture is depicted in Fig. \ref{fig:se}(b). The traversal of the decision tree is simplified to the traversal of the distribution register file. In each clock cycle, \rev{the sampling unit reads the matrix in column-wise fashion to perform the distance computing, via the port $SU.B$ of the RF}. A random number of the LFSR is combined with the column of the M-matrix is used to compute the distance $d$, following the formula in Fig.\ref{fig:se}(a). The computation logic can derive the sampled label with a first-negative value. An example of the signal waveform is shown in Fig.~\ref{fig:se}(c) when random bits are $00$. 

\textbf{Decoder.}
\rev{The decoder architecture is shown in Fig.~}\ref{fig:se}(c).  \rev{It uses the distance $d$ and corresponding matrix column $n[i]$ as input to decode the sampler output. A parallel prefix adder, comparators, and an adder tree are used to support dynamic precision.}
More details on the KY architecture can be found in \cite{sz_date}. 

\textbf{Discusion.}
With the proposed rejection sampler, we obtain throughput and accuracy benefits over \cite{sz_date} by also supporting fractional probability numbers that do not add up to a power of two, avoiding costly normalization. For a distribution with the bin size of $N$, it can reduce the runtime complexity from $O(N)$ \cite{pgma, mrf_uq} (linear search CDF sampler) and $O(log(N))$ \cite{coopmc} (binary search CDF sampler) to $O(H)$, where $H$ is the entropy of the input distribution. Moreover, the implemented sampler, nominally targeting distributions with up to 32 bins, can sample from 2, resp. 4, 8, and 16 distributions in parallel when sampling from distributions with $\leq$16, resp $\leq$8, $\leq$4, and 2 bins simultaneously. The resulting performance is summarized in table \ref{tab:ky_cdf} and benchmarked against a traditional CDF sampler \cite{lp_mcmc} and KY  sampler with rejection feature \cite{sz_date} demonstrating area, energy, and throughput benefits. 
Moreover, for energy-based models, the entropy of the input distribution keeps converging to a lower level during the PM inference, which means our sampler can speed up after a certain warm-up period on the same PM.

\subsection{Interpolation unit} \label{sec:lut_unit}
\begin{figure}[!t]
\centering
    \includegraphics[trim={0cm 0cm 0cm 0cm} , clip, width=0.82\columnwidth]{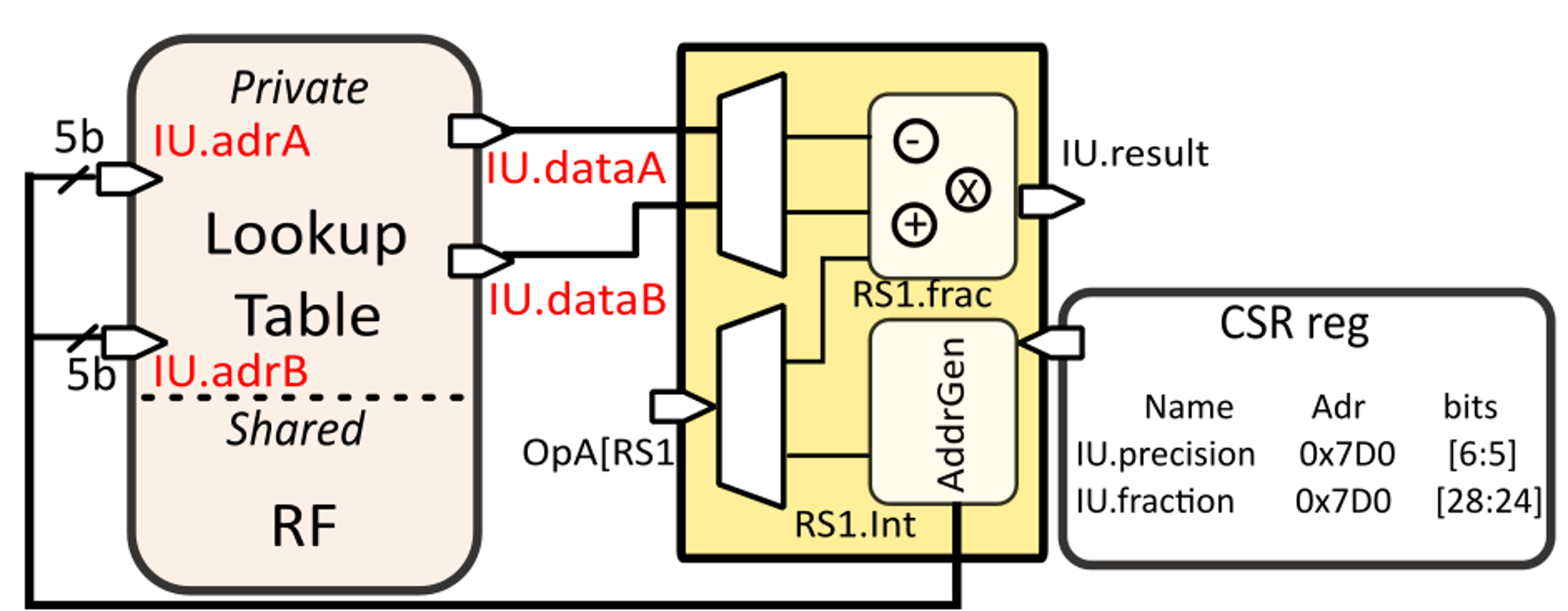}
    \caption{Interpolation unit to simplify the exp operation.}
\label{fig:lut_unit}
\end{figure}

\begin{table}[!t]
\centering
\caption{\rev{Instruction count comparison between hardware-supported and software-based LUT interpolation}}
\begin{tabular}{lcc}
\toprule
Instruction & This work & Memory-base LUT \\ \midrule
\texttt{Xprob.IU} & 1 & - \\
shift    & - & 1 \\
add      & - & 4 \\
bit and  & - & 1 \\
mult     & - & 1 \\
load     & - & 2 \\
\bottomrule
\end{tabular}%
\label{tab:lut}
\end{table}

 \rev{The interpolation unit is introduced to support efficient lookup table-based (LUT) nonlinear function calculation for the distribution computation. To simplify the hardware implementation, a uniform linear interpolation unit based on LUT entries was designed}, whose architecture is shown in Fig. \ref{fig:lut_unit}. 
 
 For our workloads, MCMC requires repetitive nonlinear operations (exp, log, etc.) in the distribution generation stage of all RVs (as shown in Eqn. (\ref{eq:bn}) and (\ref{eq:mrf})). The single-cycle hardware module 
for a lookup table with linear interpolation consists of the following components:

\textbf{Lookup Table in Private RF}.
The lookup table is stored in the private RF. Let $Y$ denote the lookup table, where $Y[i]$ represents the $i$-th entry. As each word contains 32 bits, $Y[i]$ can represent either a single full-precision entry, two 16-bit entries, or four 8-bit entries for parallel operation. 
To support the interpolation block, the RF is equipped with two additional RF read ports, $IU.adrA$ and $IU.adrB$, directly accessible in the EX stage, as shown in Fig. 3(a). 

\textbf{Address Generation Unit}.
the $IU.adrA$ and $IU.adrB$ represent the addresses of the $Y[i]$ and $Y[i+1]$ values in between the interpolation should take place. The index $i$ is computed according to the integer part of the input data $RS1$. It hence generates $IU.adrA$ and $IU.adrB$ as $\lfloor RS1 \rfloor$ and $\lceil RS1 \rceil$ in which the fractional point of $RS1$ is set through the CSR interface of the Interpolation unit.

\textbf{Interpolation Logic.} 
Previously discussed infrastructure now allows the interpolation unit to in a single clock cycle fetch the interpolation value $RS1$, as well as the Y-values of the two neighboring points 
$Y[RS1.int]$ and $Y[RS1.int+1]$ from the lookup \rev{table entries in the RF. The interpolated output $y$ can be computed as}:
\[y = Y[RS1.int] + RS1.frac\times(Y[RS1.int+1] - Y[RS1.int]).\]

\textbf{Throughput Performance.}
\rev{To finish this LUT operation, traditional software-based LUT interpolation requires 9 separate instructions. As shown in Table}~\ref{tab:lut}, \rev{2 memory loading instructions are required to load the LUT entries, together with a series of computing instructions for the address generation and result computation}. 
\rev{In our design, one single cycle instruction,} \texttt{Xprob.IU}, suffices to trigger the interpolation unit to take the input value, generate two addresses, access the lookup table memory to obtain the interpolation bounds, calculate the fractional part $f_p$, and compute the interpolated output $y$ as described above.

\textbf{Accuracy Impact.}
\rev{The accuracy of the system is influenced by the size and bitwidth of the LUT. CoopMC}~\cite{coopmc}  \rev{conducted a series of end-to-end experiments and demonstrated that, for the sampling workloads, a LUT size of 16 with a bitwidth of 8 provides a sufficient balance between accuracy and efficiency. We followed this setup in our implementation.}

\subsection{Memory design}\label{sec:local_scratchpad}
\textbf{Local scratchpad}.
Each AIA core uses a scratchpad instead of a cache for the following reasons:   
\begin{itemize}
    \item With the predefined structure of PMs, a software-managed scratchpad can selectively store only the data that has local reuse.
    \item The typical cache line granularity of 32 or 64 words is too large for PMs due to frequent irregular memory fetches for labels and CPTs, and results in wasted interconnect traffic \cite{coopmc}.
    \item A memory-mapped scratchpad avoids tag storage and lookup, reducing the area and energy footprint.
\end{itemize}

\begin{figure*}[ht]
    \centering
    \includegraphics[trim={0cm 0cm 0cm 0cm} , clip, width=0.85\textwidth]{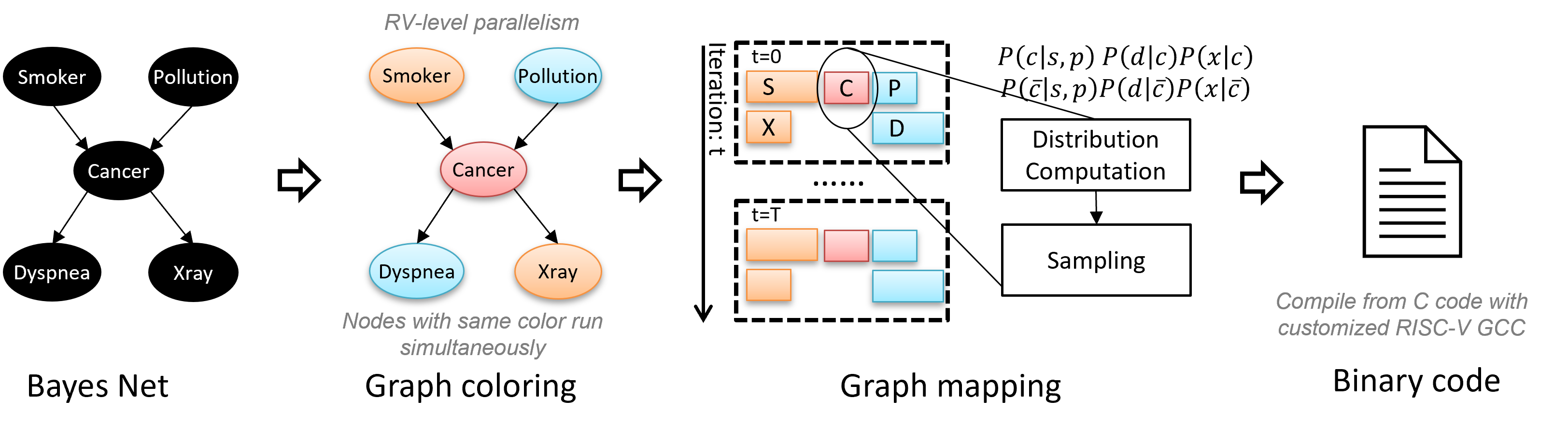}
    \caption{Customized compiler chain for AIA: consists of graph coloring to detect RV-level parallelism, graph mapping to map the RVs on different cores, and customized RISC-V GCC compiler to generate the binary code. \rev{For the example depicted here, only two cores are used for the acceleration.}}
    \label{fig:compiler}
\end{figure*}

\begin{figure*}[ht]
    \centering
    \includegraphics[trim={0cm 0cm 0cm 0cm} , clip, width=\textwidth]{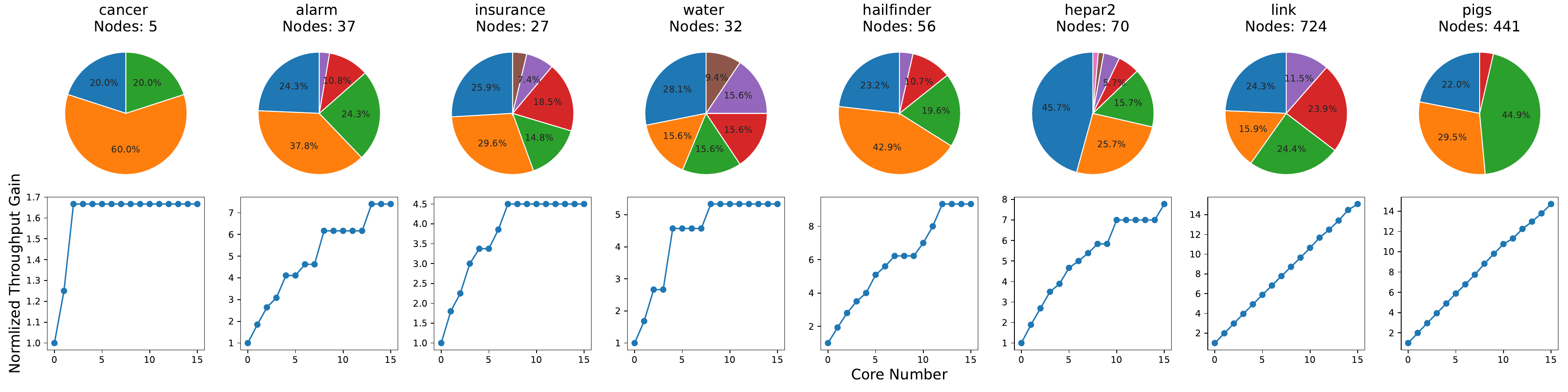}
    \caption{\rev{Graph coloring result for irregular Bayesian Networks. The Pie graph shows how many nodes have the same color. The line graph indicates the potential throughput gain on different sizes of parallel cores.}}
    \label{fig:pie_result}
\end{figure*}

\subsection{Synchronization} \label{sec:event_unit}
The single-cycle synchronization of accelerator cores is made possible with a \textit{global barrier} instruction and a global sync unit.
To mitigate the overhead of frequent synchronizations, the accelerator cores are equipped with special instructions for global barriers, complemented with a central dedicated global sync hardware unit: event unit. When a  reaches a global barrier instruction, it indicates this to the global sync unit and stalls until the other accelerator cores hit their global barrier instruction. The clock is gated during the stall to reduce power consumption. The event unit uses a tree of AND gates to determine if all the accelerator cores have reached the barrier and communicates this to all the accelerator cores within that cycle, enabling a single-cycle synchronization. Even though the paths to/from the global sync units create a long combinational loop due to the unit's centralized role, they are not critical paths in our design. 

\section{Customized Compiler} \label{sec:compiler}
A custom compiler chain  (Fig. \ref{fig:compiler}) is developed for AIA to enable rapid mapping of MCMC algorithms for a wide variety of PMs.
In our work, we used 32-bit fixed-point computing with 1-bit signed, 8-bit integer, and 23-bit fraction, following in the low-precision quantization configuration in \cite{lp_mcmc, pgma}, which shown with negligible accuracy loss.

The PMs can be described using a probabilistic programming language (PPL), aGrums~\cite{agrum}, which can convert both directed graphs and undirected graphs to factor graphs with a junction tree algorithm \cite{koski2012review}. We develop the graph coloring algorithm to split all RVs into sets of conditionally independent variables (colors), which can be updated in parallel. E.g., for an MRF, block Gibbs sampling requires only two colors in a checkerboard pattern~\cite{pgma, spu}. 
Irregular workloads, such as Bayes Nets \cite{bn_finance, bn_medicine}, typically require more color sets.

\subsection{Graph coloring}
Graph coloring is well-known as a classic NP-hard problem \cite{aslan2016performance}. In our work, the DSATUR (Degree of Saturation) graph coloring algorithm \cite{aslan2016performance} is used as a heuristic approach for coloring the vertices of a graph efficiently. The algorithm starts by initializing an empty set of colors and a set of uncolored vertices equal to the entire vertex set of the graph. In each iteration, the algorithm identifies the vertex with the maximum degree of saturation among the uncolored vertices. This is done by examining the neighbors of each uncolored vertex and counting the number of different colors they have been assigned. After selecting the vertex with the highest degree of saturation, the algorithm determines the set of colors used by its neighbors. It then selects the smallest unused color and assigns it to the selected vertex. The vertex is then removed from the set of uncolored vertices, and the color is added to the set of used colors. This process continues until there are no more uncolored vertices. The result is a coloring of the graph where no two adjacent vertices have the same color.
We checked the conditional independence properties after graph coloring, this heuristic algorithm works well for our workloads.

\rev{The coloring results for a wide range of irregular graph workloads are shown in Fig.~}\ref{fig:pie_result}. \rev{The pie charts illustrate the ratio of nodes assigned to each color, showing that the number of used colors never exceeds six. This indicates that our graph coloring algorithm produces a relatively balanced distribution. The line graphs underneath depict the achievable throughput gains across varying numbers of parallel processing cores. Smaller graph models, such as 'cancer', show fewer benefits from scaling up the core count, which is caused by the limited random-variable level parallelism inherently present in these small graphs. 
Most workloads, on the other hand, demonstrate excellent scalability using our approach for large, sparse graph models.}

\subsection{Graph mapping and binary code generation}
In the subsequent step, mutually independent nodes are mapped onto the 16 parallel accelerator cores, using the information from the graph coloring stage, with a heuristic that maximizes the parallelism and minimizes the communication distance between nodes that have to exchange information. Finally, using the RISC-V GCC compiler, enhanced with our custom instructions, a binary executable is generated for each RISC-V core to execute the respective probability generation and node sampling.

\section{Experiments} \label{sec:experiments}


\subsection{Area, frequency and power}
AIA is taped-out in Intel 16 nm technology with an area of 2.0$\times$2.0 mm$^2$, where Fig. \ref{fig:chip_micrograph} shows the layout of the SoC and the packaged chip. 
The design was synthesized with Cadence Genus 2021.10, and Place \& Route was performed with Innovus 2021.11. Synopsys ICV 2022.12 is used for DRC. LVS was checked with Mentor Calibre 2022.3. The Hammer flow \cite{hammer} is used to manage the usage of different EDA tools in different steps. One single power domain with the normal supply of $0.8 V$ is used for the main core of the chip. IO and LDO of PLL use the supply of $1.2 V$.

\begin{figure}[!t]
    \centering
    \includegraphics[trim={0cm 0cm 0cm 0cm} , clip, width=0.82\columnwidth]{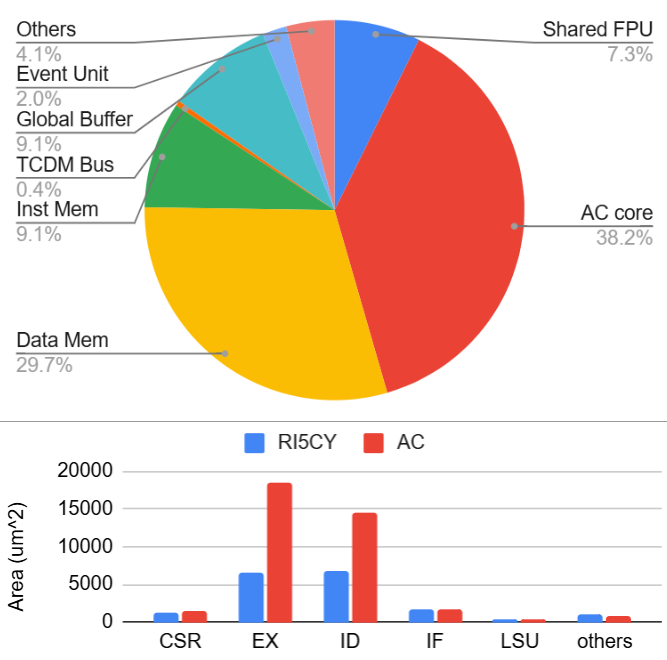}
    \caption{(a) Area percentage of the MESH domain. (b) Area breakdown comparison with \rev{RI5CY} \cite{pulp} baseline design.}
    \label{fig:area}
\end{figure}

Fig.~\ref{fig:area} summarizes the post-synthesis area breakdown of the chip. The $MESH$ domain occupies an area of 1.5 mm$^2$. The accelerator cores with the data and instruction memory occupy almost $70\%$ of the $MESH$ area. 
For further insight, Fig. \ref{fig:area}(b) compares the area of the different pipeline stages of the accelerator core to the baseline design \cite{marsellus}. The area of the Instruction Decode (ID) stage has doubled, primarily due to the increase from a 32-word to a 64-word register file. The area of the Execution (EX) stage has also increased, attributed to the addition of the sampler unit and interpolation unit. Other modules, such as the Control Status Registers (CSR), Instruction Fetch (IF), and Load-Store Unit, have maintained a similar area.

%

\begin{figure}[!t]
\centering
\includegraphics[trim={0cm 0.4cm 0cm 0cm} , clip, width=\columnwidth]{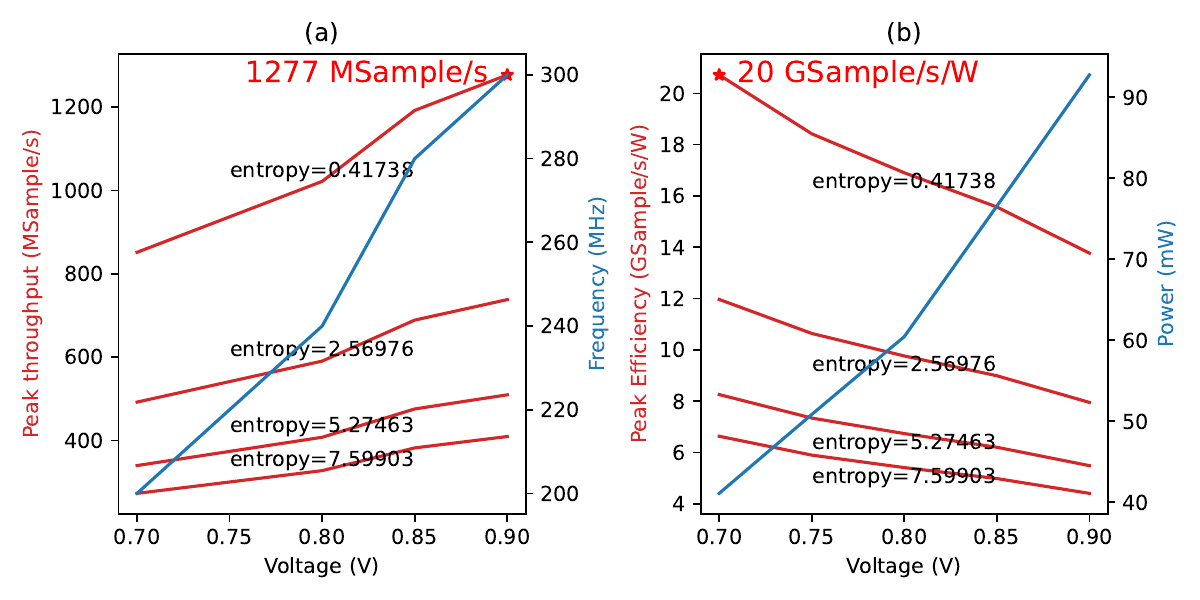}
\caption{Peak-performance scaling with voltage and entropy.}%
\label{fig:voltage_scaling}
\end{figure}%

Fig. \ref{fig:voltage_scaling} shows the measured SoC performance when running a typical sampling from the distribution workload. The chip's electrical performance is measured by scaling the voltage from the nominal 0.9V to 0.7V. Our chip can operate up to the maximum frequency of 300 MHz at 0.9V.  As discussed in section \ref{sec:se_unit}, the sampling performance of our chip depends on the entropy $H$ of the distribution. For the distributions with the lowest entropy, it has a peak throughput of 1277 MSamples/s (Fig. \ref{fig:voltage_scaling}(a)) and a peak energy efficiency of 20 GSample/s/W at 0.7V (Fig. \ref{fig:voltage_scaling}(b)). Thanks to the custom sampler accelerators in the execution stage, the mesh can execute up to 16 Samples/Cycle on each of the sixteen cores, \rev{which introduces a peak performance of 256 samples/cycle}.

 \subsection{End-to-end workloads}\label{sec:workloads}
The performance of the chip is also benchmarked with complete end-to-end PM inference workloads, again representing the two different types of PMs: Bayes Nets and MRF. \rev{Following the experiment setup in Statheros}~\cite{lp_mcmc} and MSSE~\cite{pgma}, \rev{we used 32b fixed point computing with 1b signed, 7b integer, and 24b fractional bits in our experiment. The data format can be further optimized, which is beyond the scope of this paper.} The experiments are performed with PMs that fit entirely in the on-chip data memory. The programming of the instruction memory and data memory is done through an FPGA via a slow IO interface, which is not included in the performance results.

\begin{figure}[!t]
    \centering
    {\includegraphics[trim={0cm 0cm 0cm 0cm} , clip, width=\columnwidth]{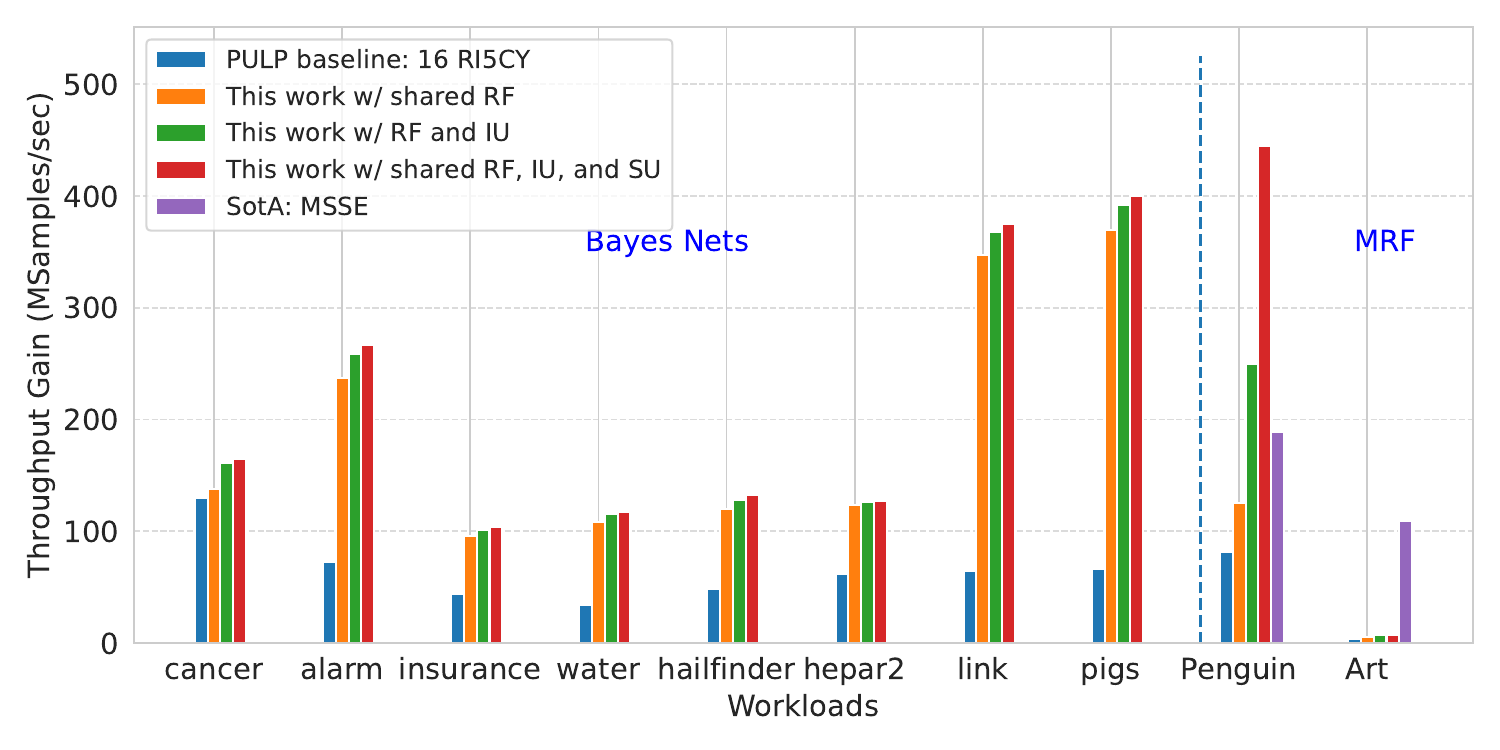}}
    \caption{\rev{Throughput performance gains over baseline design and introduced hardware modules.}}
    \label{fig:throughput}
\end{figure}

\textbf{irregular PMs}.
We used Bayes Nets as the benchmarks to compare AIA to different hardware platforms. Bayes Nets are used in machine learning for healthcare, robotics, and safety-critical tasks \cite{bn_finance, bn_medicine, bn_xai}. The experiments are run on the standard benchmarks for Bayes Nets from the BNrepository \footnote{\url{https://www.bnlearn.com/bnrepository/}}.
These workloads from the BN repository are suitable for performance evaluation due to their diversity in size and structure. The runtime results for the single marginal probability across different benchmarks are summarized in Table \ref{table:runtime} and Fig. \ref{fig:throughput}.

\textbf{Regular PMs}.
We use MRFs as a representative workload for regular graph PM inference in our benchmarking section. MRFs can be used for a broad range of applications, such as image denoising, stereo matching, speech recognition, etc. \cite{mrf_uq, pgma}. Following previous work, two MRF workloads \cite{zhang2021statistical}, Penguin and Art, are mapped onto our chip. 
In Fig. \ref{fig:throughput} and Table IV, AIA's performance for these \textit{regular} PMs is compared with PULP as well as the SotA ASIC design, MSSE \cite{pgma}, for MRF inference. 
It is important to note that the existing PM inference accelerator ASICs can only map MRFs, i.e. 2D-grid mesh workloads. 

\begin{figure}[!t]
    \centering
    \includegraphics[trim={0cm 0cm 0cm 0cm} , clip, width=\columnwidth]{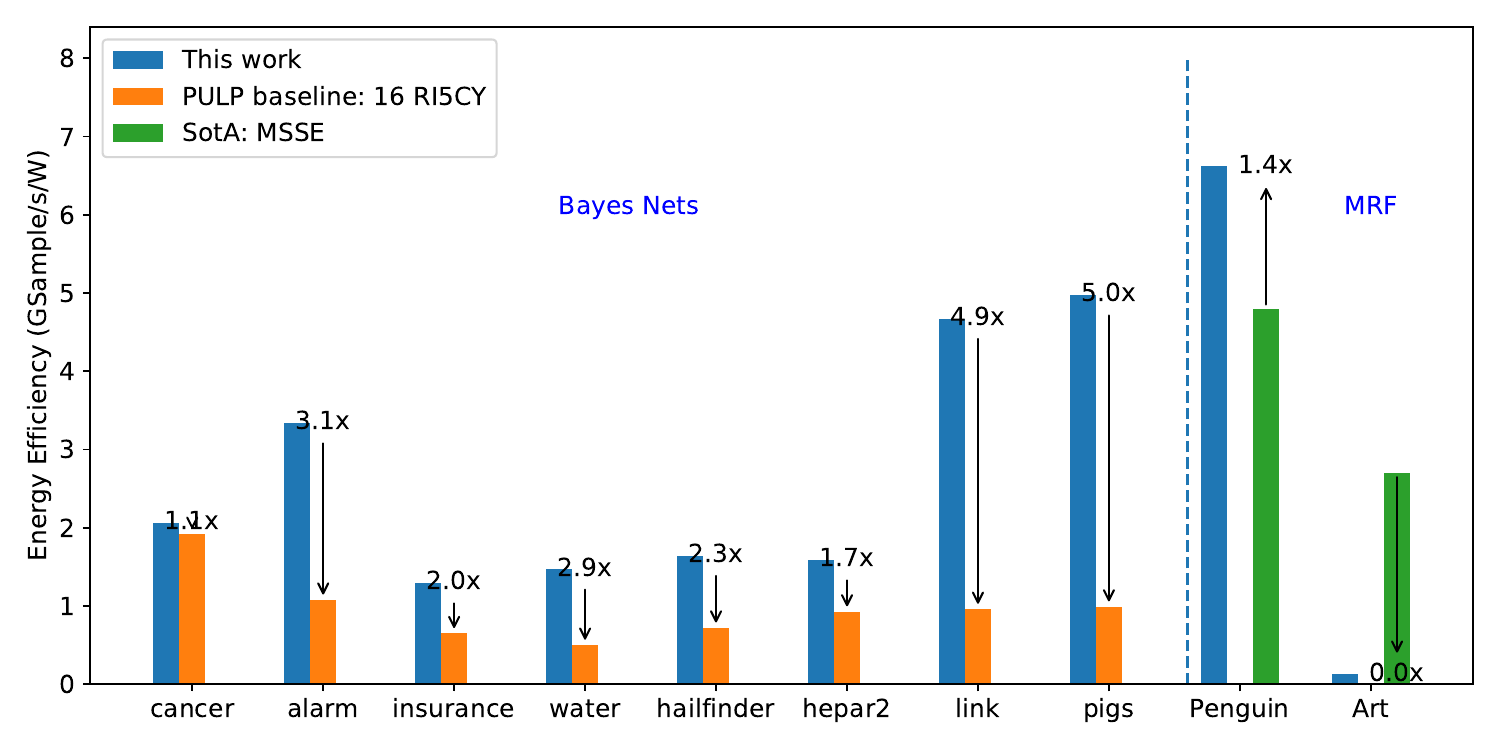}
    \caption{Throughput and energy performance compassion on real-world applications.}
    \label{fig:energy}
\end{figure}

\begin{figure}[!t]
\centering
\includegraphics[trim={0cm 0cm 0cm 0cm} , clip, width=0.9\columnwidth]{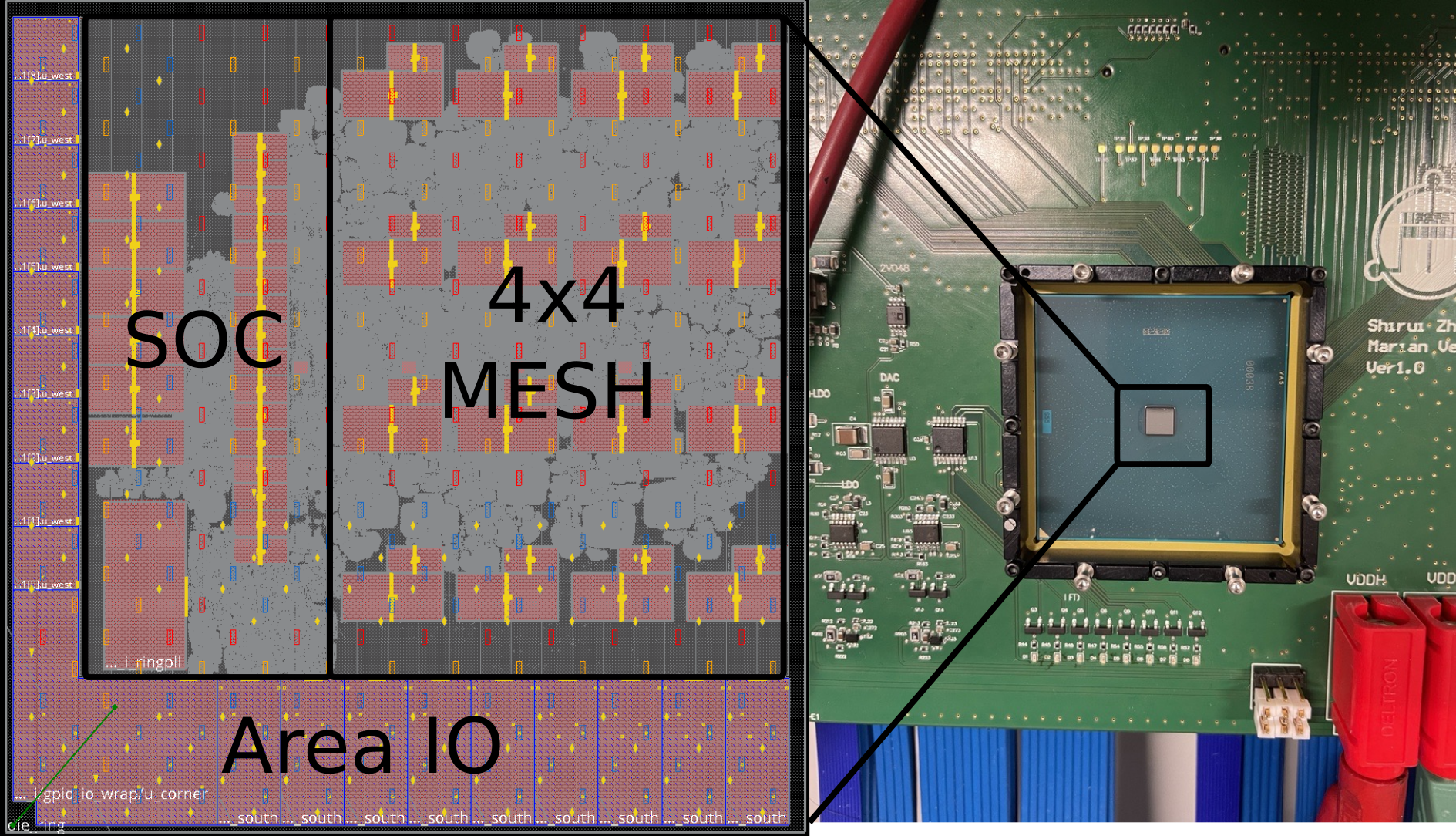}
\caption{Chip layout of post routing and the package chip.}%
\label{fig:chip_micrograph}
\end{figure}

\textbf{PULP baseline.} 
We compared the throughput and energy performance when deploying the workload on the parallel CPU baseline via power simulations. In the PULP baseline \cite{pulp}, there's no such custom sampler, interpolation unit, and enlarged RF. The result is plotted in Fig. \ref{fig:throughput} and Fig. \ref{fig:energy}. \rev{For a fair comparison, the clock speed of the PULP platform was set to 250MHz with 16 parallel cores. Moreover, as the KY sampler can not be implemented efficiently in software, a CDF sampling algorithm was used with minimum normalization to maximize PULP's performance.} Compared to this baseline design, AIA achieved an average improvement of 2$\times$ in terms of throughput and 1.45$\times$ in terms of energy efficiency.
\textbf{Throughput gain breakdown.}
\rev{We further study the impact of each proposed innovation on the total system throughput performance shown in Fig.~}\ref{fig:throughput}. \rev{For Bays nets, compared to small graph models, such as cancer, insurance, etc., large models show higher throughput gain, due to the nature of graph coloring-based parallel MCMC described in Sec.~}\ref{sec:compiler}. \rev{The enlarged RF with inter-core sharing contributes the most of the gain in Bayes nets because we can leverage the large RF to avoid frequent memory access for intermedia data and global buffer access. In the Penguin MRF workload, the sampling unit contributes more throughput gain as sampling takes most of the runtime for this workload. These results align with the runtime breakdown result shown in Fig.~}\ref{fig:op_breakdown}(a). \rev{Sampling dominate workloads gain more for hardware sampler, while the other workloads gain for enlarged RF and inter-core RF sharing.}


\textbf{CPU/GPU performance comparison.} 
We used three different software implementations to show the performance on general-purpose hardware such as CPUs and GPUs, running various exact as well as approximate inference algorithms.
\begin{itemize}
    \item Dice \cite{dice} is a SotA exact inference PPL framework based on CPU targeting fast exact inference for discrete probabilistic programs. 
    \item pyAgrum \cite{agrum} is a scientific library dedicated to performing efficient computations with Bayes Nets and other probabilistic graphical models, which show promising approximate inference efficiency on the CPU.
    \item Bayeslib \cite{Bayeslib} 
    is an open-sourced C++-based parallel inference library for discrete Bayes Nets that supports approximate inference algorithms both in CPU and GPU.
\end{itemize}

    \begin{table}[!b]
    \centering
    \caption{Single marginal runtime comparison results (unit: millisecond)}
    \label{table:runtime}
    \begin{threeparttable}
    \begin{tabular}{c |c c c c | c }
    \toprule

      { Benchmark}  & { Dice \cite{dice}}  & {pyAgrum \cite{agrum}} & \multicolumn{2}{c|}{Bayeslib \cite{Bayeslib}}    & \textbf{Ours} \\ 
    \midrule
    \textbf{Platform} & CPU & CPU & CPU & GPU & ASIC\\
    \midrule
    survey  & 810 & 10 & 1130 & 1681 & \textbf{0.5}\\  
    cancer  & 690 & 10 & 720 & 1134  & \textbf{0.5} \\
    alarm   & 5730 & 120 &  14130 &  49576 & \textbf{1.8} \\    
    insurance  & 4620 &100 & 19790 &  40625 & \textbf{3.4} \\
    water & 282790 & 140 & 40  & $\ddagger$ & \textbf{2.2}\\
    hailfinder  & 22280 & 210 &  157331& 198614 & \textbf{7.5}\\
    hepar2  & 11620 & 390 & 40280 & 154518 & \textbf{4}\\
    pigs  & 136790 & 9150 & 43 & $\ddagger$ &\textbf{16.2}\\
   
    \bottomrule
    \end{tabular}
        \begin{tablenotes}
        \footnotesize
        \item[1] $\ddagger$ \textit{the program throwing runtime error log}
        \end{tablenotes}
    
    \end{threeparttable}
\end{table}

All results are summarized in Table \ref{table:runtime}. The performance of AIA's sampling-based inference algorithms is first of all clearly better compared to the exact inference of Dice in terms of scalability and generality. 
While a well-developed approximate inference platform, like pyAgrum \cite{agrum}, can handle it within one minute. For exact inference on CPU, the runtime depends both on the model size and the graph structure, while approximate inference only depends on the model size. 
Another benefit is that during the sampling procedure, it can compute all the single marginal distributions without introducing computation overhead if there's enough memory to store the histogram.

Compared to the CPU-based sampling-based inference algorithms, we achieved an average speedup of 60x over the Bayes Nets workloads. The reason is that although the CPU has a very high single-core performance with a GHz clock frequency, it fails to exploit the parallelism opportunities existing in PM graphs due to the frequent memory data exchange. 

Baylib \cite{Bayeslib} is the only open-sourced Gibbs sampling framework developed based on GPU. It aims to utilize the massive parallelism of modern parallel computing platforms to boost the sampling performance. But as shown in the runtime table \ref{table:runtime}, the throughput isn't improved because GPU synchronization adds a large cost to synchronize all the samples, which is frequently needed in these workloads. It also fails to map the workloads with complex graph structures, with reported errors when mapping the $water$ and $pigs$ models on GPU.

\textbf{SotA ASIC and FPGA}.
Fig. \ref{fig:throughput} also shows that the SotA accelerator surpasses AIA in performance on this specific workload, Art.
The main reasons are that this SotA is heavily optimized for a two-color algorithm (typical for MRF) and contains more dedicated functional units to efficiently (yet inflexible) compute the different special functions 
(e.g., exp, log). Our design instead favored flexibility and programmability at the expense of workload performance throughput.

A last related hardware design for irregular graph PM inference is based on FPGA \cite{acmc2}. They applied the same RV- and computational-level parallelisms on a Network-on-Chip (NoC) based design as in AIA. However, this work requires generating different FPGA bitstreams for different workloads and suffers from the large power consumption of FPGAs, being less efficient than ASICs (see Table \ref{table:sota}).

\begin{table}[!b]
\centering
\caption{Comparison table with SotA}
    \label{table:sota}
\begin{tabular}{lccccc}
\toprule
 & This work & MSSE~\cite{pgma} & AcMC$^2$~\cite{acmc2} \\ 
 \midrule
Technology & 16nm & 16nm & FPGA  \\
Die Area & 4mm$^2$ & 5mm$^2$ & - \\
SRAM  & 960KB & 103KB & 4MB \\
Sampler No\# & 16 & 12 & 16  \\
$F_{max}$ & 300MHz & 651MHz & 163MHz   \\
Power@$F_{max}$ & 93.5mW@0.9V & 42.2mW@0.8V & 16.3W   \\
\makecell[l]{Peak Throughput \\ (GSample/s)} & 1.27@0.9V & 0.372$*$ & 2.6  \\
\makecell[l]{Peak Energy Effi. \\ (GSample/s/W)} & 20@0.7V & 17.6$*$ & 0.16  \\
\makecell[l]{Peak Area Effi. \\ (GSample/s/mm$^2$)} & 0.567@0.9V & 0.284$*$ & -  \\
\midrule
Sampler & KY & CDF & \makecell[c]{Alias-table\\ RNG }  \\ \midrule
Model & \makecell[c]{\textbf{General PMs} \\ (MRF, BayesNet)} & \makecell[c]{Only\\MRF} & \makecell[c]{General\\ PMs}   \\ \midrule
Inference & \makecell[c]{\textbf{discrete MCMC} \\ (Gibbs, MH, etc.)} & \makecell[c]{Only\\Gibbs} & \makecell[c]{MCMC}   \\
\bottomrule
\end{tabular}%
\begin{tablenotes}
    \footnotesize
    \item $*$: contains both sampling and computing
\end{tablenotes}
\end{table}


\section{Conclusion} \label{sec:conclusion}
This work presents a 4mm$^2$ multicore SoC in 16nm, which executes approximate inference acceleration of PMs. It contains 4x4 customized RISC-V cores, where each core features a non-normalized KY sampler and interpolation unit, as well as core-to-core direct data access via the register file. A customized compiler chain is developed to map MCMC algorithms for a wide variety of PMs. The results and wide variety of workloads mapped demonstrate the performance gains and flexibility compared to the SotA implementations.

\section*{Acknowledgment}
This project has been partly funded by the European Research Council (ERC) under grant agreement No. 101088865, the Flanders AI Research Program, Research Foundation-Flanders (FWO) under grant 1SE7723N, and KU Leuven.


%



\clearpage

\ifCLASSOPTIONcaptionsoff
  \newpage
\fi



\bibliographystyle{IEEEtran}
\bibliography{refs.bib}

\end{document}